\documentclass[12pt,prl,twocolumn,floatfix,showpacs,
superscriptaddress]{revtex4}
\usepackage{fullpage}
\usepackage{graphicx}
\usepackage{amsmath2000}
\usepackage{citesort}

\newcommand{\E}{\varepsilon}
\newcommand{\be}{\begin{equation}}
\newcommand{\ee}{\end{equation}}
\newcommand{\bea}{\begin{eqnarray}}
\newcommand{\eea}{\end{eqnarray}}
\newcommand{\beas}{\begin{eqnarray*}}
\newcommand{\eeas}{\end{eqnarray*}}
\newcommand{\nnp}{\nonumber\\}
\newcommand{\ve}[1]{{\bf #1}}
\newcommand{\mus}{\mu^\star}
\newcommand{\lbr}{\left [}
\newcommand{\rbr}{\right ]}
\newcommand{\lb}{\left \{}
\newcommand{\rb}{\right \}}
\newcommand{\lp}{\left (}
\newcommand{\rp}{\right )}
\newcommand{\ld}{\left .}
\newcommand{\rd}{\right .}

\newcommand{\sli}{\sum\limits}
\newcommand{\ili}{\int\limits}

\newcommand{\atof}[1]{\alpha^2_{#1}F(\omega)}
\newcommand{\itof}[1]{I^2_{#1}\chi(\omega)}
\begin{document}
\title{Application of an Extended Eliashberg Theory to
High-$T_c$ Cuprates}
\author{E. Schachinger}
\email{schachinger@itp.tu-graz.ac.at}
\homepage{www.itp.tu-graz.ac.at/~ewald}
\affiliation{Institut f\"ur Theoretische Physik, Technische Universit\"at
Graz\\A-8010 Graz, Austria}
\author{J.P. Carbotte}
\affiliation{Department of Physics \& Astronomy, McMaster University\\
 Hamilton, Ontario, Canada L8S 4M1}
\begin{abstract}
In recent years a unified phenomenological picture for
the hole doped high-$T_c$ cuprates has emerged for a spin and charge
spectroscopy. Spectral anomalies have been interpreted as
evidence of charge carrier coupling to a collective spin
excitation present in the optical conductivity,
in ARPES (angular resolved photoemission), and in
tunneling data. These anomalies can be used to derive an
approximate picture of a charge carrier-exchange boson interaction
spectral density $\itof{}$ which is then be used within an
extended Eliashberg formalism to analyze normal and
superconducting properties of optimally doped and overdoped cuprates.
This paper reviews recent developments and demonstrates the
sometimes astonishing agreement between experiment and
theoretical prediction.
\end{abstract}
\maketitle

\section{Introduction}

The standard Eliashberg equations \cite{Eli,Carb1}
were derived for superconductors
with an energy gap of $s$-wave symmetry and the electron-phonon
interaction as the pairing interaction. This type of
interaction allows the application of Migdal's theorem which
states that vertex corrections in the electron-phonon interaction
can be neglected to order $\omega_D/\E_F$, with $\omega_D$ the
Debye energy and $\E_F$ the Fermi energy. On the other hand, it is
now widely accepted that the high $T_c$ cuprates have an energy
gap of $d_{x^2-y^2}$ symmetry \cite{Hardy,Bonn,ZShen,Wollm,Tsuei,Broun}
and there is still no consensus as to
the microscopic mechanism leading to Cooper pairs in these materials.

In Eliashberg theory a given superconductor is characterized
by the Eliashberg function $\atof{}$ which describes the exchange
of a phonon by two
electrons at the Fermi surface and by the Coulomb potential
$\mus$. These are the kernels in the two non-linear coupled
Eliashberg equations. One equation, which is referred to as
the renormalization channel, describes the effect of the
electron-phonon interaction on normal-state properties
modified further by the onset of superconductivity. The
second equation, referred to as the pairing channel, deals
with the energy gap directly and is identically zero in
the normal-state. When reliable tunneling data is available
for the quasiparticle density-of-states $N_{qp}(\omega)$,
for instance, the
procedure can be inverted \cite{BCS6,McMillan} to get
from $N_{qp}(\omega)$ the kernels $\atof{}$ and $\mus$.
In principle, it should also be possible to get the same
information from infrared data \cite{Joyce,Allen1,Farnw}
although in conventional systems this has not been widely
done while tunneling has. Once the kernels $\atof{}$ and
$\mus$ are known, the finite temperature Eliashberg
equations can be solved numerically to obtain superconducting
properties.

In principle, the Eliashberg equations can easily be generalized to
include $d$-wave symmetry of the energy gap by an appropriate
extension (to include a dependence on orientation of the electron
momenta) of the charge carrier-exchange boson interaction
spectral density ($\alpha^2F(\omega)$ in
case of the electron-phonon interaction) which contains all the
relevant information about the coupling of the charge carriers
to the exchange bosons. As the microscopic mechanism leading to
superconductivity is not yet known, information on the charge
carrier-exchange boson interaction spectral density (denoted
$I^2\chi(\omega)$ throughout this paper) is to be obtained by
a fit to appropriate data sets using phenomenological models.
Such a procedure can yield a first approximation to a
complete description even in cases when an equivalent to Migdal's
theorem is not applicable and vertex corrections are not
entirely negligible.

Unfortunately, the well established inversion techniques
which allowed one to determine $\alpha^2F(\omega)$ from
tunneling experiments \cite{McMillan} have, so far, not been extended
to the cuprates and, therefore, phenomenological models had
to be developed for $I^2\chi(\omega)$.
One such phenomenological model has been introduced by
Schachinger {\it et al.} \cite{schach1,schach2,schach3} and
was reviewed by Schachinger and Sch\"urrer \cite{SchachRev}.
This model is a purely electronic model and describes
the feedback effect the superconducting state has on
$I^2\chi(\omega)$. The authors used, for definiteness, the
spin fluctuation model introduced by Pines and coworkers
\cite{Pines1,Pines2} in their Nearly Antiferromagnetic
Fermi Liquid (NAFFL) model. The feedback effect caused by
superconductivity is described by introducing a low energy gap in
$I^2\chi(\omega)$ (low frequency cutoff) which opens up
as the temperature is lowered through the critical temperature
$T_c$. This gap shows the same temperature dependence and
size as the superconducting energy gap. Within this model
it was possible to describe consistently the temperature
dependence of the microwave conductivity with its pronounced
peak around $40\,$K observed in optimally doped
YBa$_2$Cu$_3$O$_{6.95}$ (YBCO) \cite{BonnA,Srikanth}, the
similar peak observed in the electronic thermal conductivity
\cite{Matsukawa}, and the temperature dependence of the
penetration depth in nominally pure YBCO samples and in
YBCO samples with Zn or Ni impurities \cite{schach1}.
Nevertheless, similar results would have been achieved
using the Marginal Fermi Liquid (MFL) model \cite{Varma}
assuming a $d$-wave gap together with a low frequency cutoff
to describe the charge carrier-exchange boson interaction
spectral density.

A remarkable step forward in the development of phenomenological
models was provided by the work of Marsiglio {\it et al.}
\cite{mars4} who were able to show analytically that there
exists a simple, approximate formula which relates
$\alpha^2F(\omega)$ to the normal state optical conductivity
$\sigma(\omega)$ via the second derivative of the real part
of $\omega\sigma^{-1}(\omega)$. This established the basis
for a spectroscopy which allows the measurement of the spectral
density $\alpha^2F(\omega)$ directly from optical data. This
result was then extended to the superconducting state of
$d$-wave superconductors by Carbotte {\it et al.} \cite{schach4}
who explore the relationship between spectral density and
$W(\omega)$ at low temperatures in the superconducting state.
They conclude that the relationship is not at all as direct,
but, even though more complicated, it remains simple enough
to be very useful although more approximate.
(A similar procedure was also suggested by Munzar {\it et al.}
\cite{Munzar}.)

It will be the purpose of this paper to review the application
of this technique to various cuprates in some detail.
Thus, the paper establishes in section two the formalism,
section three discusses its application to optimally doped
YBCO. Other cuprates are also investigated within
the same context, and finally, in section four a summary
is presented.

\section{Formalism}
\subsection{The Normal State Optical Conductivity}

The optical conductivity is related to the current-current
correlation function. The paramagnetic part of the response
function on the imaginary frequency axis is given by
\cite{Mars1,lee,dolgov,shulga,schur1} 
\bea
\Pi (i\nu_n) &=& \frac{1}{N\beta} \sum_{{\bf k},m} (ev_x)^2
{\rm tr}\lb
\hat{G}({\bf k},i\omega_m)\right.\nonumber\\
  &&\times\left. \hat{G}({\bf k},i\omega_m+i\nu_n)\rb,
\label{eq:para}
\eea
where $\hat{G}({\bf k},i\omega_m)$ is a matrix Green's function
in the Nambu formalism \cite{Nambu}, $i\omega_m = i\pi T(2m+1),
m = 0,\pm 1,\pm 2,\ldots$ is the fermion and
$i\nu_n = 2in\pi T, n = 0,\pm 1,\pm 2,\ldots$ is the boson Matsubara
frequency; $T$ is the temperature and $v_x$ the component of the
electron velocity in $x$-direction. 
The factors preceding the summations include the total number of atoms
in the crystal, $N$, and the inverse temperature, $\beta \equiv 1/k_BT$.

The optical conductivity is related to the response function through
\be
  \sigma(\omega) = \frac{i}{\omega}\Pi(\omega+i0^+).
  \label{eq:sigma}
\ee
After analytical continuation to the real frequency axis and
using the usual procedure
\[
 \frac{1}{N}\sli_\ve{k}\quad\longrightarrow
 \int\!d\E\,N(\E),
\]
we arrive at a general expression
for the optical conductivity $\sigma'(\omega)$:
\bea
 \sigma'(\omega) &=& \frac{1}{i\omega}\lb\ili_{-\infty}^0\!d\nu\,
   \tanh\lp\frac{\nu+\omega}{2T}\rp \rd\nonumber\\
  &&\times S^{-1}(T,\omega,\nu)\nonumber\\
   &&+\ili_0^\infty\!d\nu\,\lbr\tanh\lp\frac{\nu+\omega}{2T}\rp\rd
   \nonumber\\
  && \ld\ld   -\tanh\lp\frac{\nu}{2T}\rp\rbr
   S^{-1}(T,\omega,\nu)\rb,
 \label{eq:42}
\eea
which has been given by Lee {\it et al.} \cite{lee} (with the factor
$ne^2/m$ suppressed). Here, $N(\E) \equiv N(\E_F)\equiv N(0)$,
$N(0)e^2 v_F^2 = \Omega_p^2/4\pi \equiv
ne^2/m$; $\Omega_p$ is the plasma frequency, $e$ the
charge on the electron, $m$ its mass, and $n$ the electron
density per unit volume. In Eq.~(\ref{eq:42})
\bea
  S(T,\omega,\nu) &=& \omega+\Sigma^\star(T,\nu+\omega)\nonumber\\
  &&- \Sigma(T,\nu)-i\pi t^+
  \label{eq:43}
\eea
with the self energy $\Sigma(T,\omega)$
related to the electron-phonon spectral density by
\bea
  \Sigma(T,\omega) &=& -\int\!dz\,\alpha^2F(z)\lbr
  \psi\lp\frac{1}{2}+i\frac{\omega+z}{2\pi T}\rp\right.\nonumber\\
  &&\left. -
  \psi\lp\frac{1}{2}+i\frac{\omega-z}{2\pi T}\rp\rbr,
  \label{eq:44}
\eea
where $\psi$ is the digamma function. In Eq.~(\ref{eq:43})
$\pi t^+ \equiv 1/(2\tau_{imp})$ gives the impurity contribution
to the electronic scattering. ($\tau_{imp}$ is the impurity
scattering time.)

\begin{figure}[t]
\includegraphics[width=8cm]{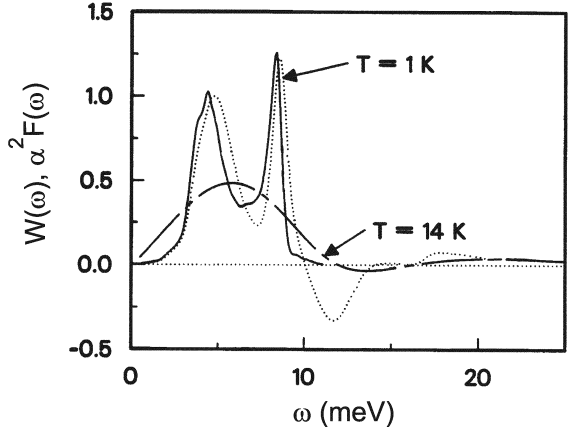}
\caption{
The solid curve is $\atof{}$ vs. $\omega$ for Pb. The other
curves are the function $W(\omega)$ according to Eq.~(\protect\ref{eq:47})
obtained from the normal-state conductivity optical scattering
rate $\tau^{-1}_{op}(\omega)$ at various temperatures.
See Marsiglio {\it et al.} \protect\cite{mars4}.
\label{fig:1}}
\end{figure}

At zero temperature these expressions for the conductivity reduce
to a simple form \cite{mars4,mars5}:
\be
  \sigma(\omega) = \frac{\Omega_p^2}{4\pi}\frac{i}{\omega}
  \ili_0^\omega\!d\nu\,\frac{1}{\omega-\Sigma(\nu)-\Sigma(\omega-\nu)},
  \label{eq:45}
\ee
and the self energy
\beas
  \Sigma(\omega) &=& \ili_0^\infty\!d\Omega\,\alpha^2F(\Omega)
  \ln\left\vert\frac{\Omega-\omega}{\Omega+\omega}\right\vert
  \nonumber\\ && -
  i\pi\ili_0^{\vert\omega\vert}\!d\Omega\,\alpha^2F(\Omega)
\eeas
for the electron-phonon interaction.
In this form Marsiglio {\it et al.} \cite{mars4} were able to
show analytically that a remarkably simple formula could be used
to establish an approximate but very useful
relationship between $\sigma(\omega)$
and $\atof{}$. The observation was also backed up by detailed
numerical work. We begin by defining an optical scattering rate
$\tau^{-1}_{op}(\omega)$ as \cite{mars4,schach4}
\be
  {1\over\tau_{op}(\omega)} = {\Omega_p^2\over 4\pi}\,
  \Re{\rm e}{1\over\sigma(\omega)} \equiv
  \Re{\rm e}{1\over\sigma'(\omega)},
  \label{eq:46}
\ee
which is routinely obtained in optical experiments. We then
define an auxiliary function \cite{mars4,schach4}
\be
  W(\omega) \equiv {1\over 2\pi}{d^2\over d\omega^2}
  \lb{\omega\over\tau_{op}(\omega)}\rb.
  \label{eq:47}
\ee
Marsiglio {\it et al.} \cite{mars4} have shown that in certain
circumstances
\be
  \alpha^2F(\omega) \simeq W(\omega),
  \label{eq:48}
\ee
which serves as a basis for a spectroscopy which allows the
measurement of the  spectral density $\atof{}$ directly from
optical data. In Fig.~\ref{fig:1} we show theoretical results
for $W(\omega)$ at two temperatures based on the case of Pb. The
solid curve is the Pb $\atof{}$ obtained from tunneling
data. The other two curves were obtained by calculating
$\sigma(\omega)$ from Eq.~(\ref{eq:42}) and computing $W(\omega)$
defined by Eq.~(\ref{eq:47}). Such calculations were performed
at two temperatures, namely $T=1\,$K (dotted) and $T=14\,$K
(dash-dotted). Within the energy range corresponding to
the range of $\atof{}$ the dotted curve for $W(\omega)$
is remarkably close to the solid curve for $\atof{}$
and therefore $W(\omega)$ gives an accurate measurement of the
absolute value as well as the frequency dependence of the
spectral density. As the temperature is increased this is no
longer the case although some rough correspondence remains
which provides a qualitative similarity between the two
quantities which could still be exploited to get a rough
first measure of the spectral density in cases where low
temperature data are not available. We point out that even
at $T=1\,$K there are negative tails in $W(\omega)$ above the
maximum phonon cutoff which are not in $\atof{}$. This
is expected since $W(\omega)$ and $\atof{}$ are not the
same quantities. In fact, it is indeed remarkable that they
should correspond so closely below the phonon cutoff energy.
This close correspondence can be exploited to get a good
first measure of the spectral density $\atof{}$ from
infrared data.
In principle, one should use a first iteration for 
$\atof{}$ obtained from the second derivative of the
conductivity defining $W(\omega)$ [Eq.~(\ref{eq:47})], to
calculate from it $\sigma(\omega)$ based on Eqs.~(\ref{eq:42}) to
(\ref{eq:44}) and keep iterating until an exact correspondence
between calculated and measured $\sigma(\omega)$ results has
been achieved. In any of the applications so far this has not
been attempted because of the many uncertainties that remain.

\subsection{The Superconducting State}

In as
much as BCS theory applies, any effective interaction between
two electrons at the Fermi surface, which is attractive, will
lead to superconductivity. This can arise from the
electron-phonon interaction through the polarization of the
system of ions. In this case we can describe the polarization
process as due to the exchange of a phonon between a pair of
charge carriers. An obvious extension is to ask: could we
exchange some other excitation? In the Nearly Antiferromagnetic
Fermi Liquid (NAFFL) model of Pines and coworkers \cite{Pines1,Pines2}
it is envisaged that spin fluctuations replace the phonons.
The basic formalism for dealing with this new situation are the
Eliashberg equations but now the $\ve{k}$, $\ve{k}'$ anisotropy in
momentum of the kernel
$\atof{\ve{k}\ve{k}'}$ needs to be taken into
account so that the resulting superconducting state exhibits
$d$-wave symmetry
to accord with the experimental observation. In the
renormalization channel we take for simplicity only the
isotropic contribution from the electron-spin fluctuation
exchange written as $\itof{}$ where $I^2$ is to denote a
spin-charge exchange coupling constant and $\chi(\omega)$
is the spin susceptibility. To get a $d$-wave gap we use
a separable interaction of the form
$\cos(2\theta)g\itof{}\cos(2\theta')$ with $\theta$ and
$\theta'$ the direction of the initial (\ve{k}) and final
$(\ve{k}')$ momentum which, for simplicity, we pin on the
Fermi surface although in the NAFFL the entire Brillouin zone
is averaged over i.e.: is not pinned on
the Fermi surface. While, for simplicity, we have assumed the
same form $\itof{}$ to hold in the pairing as in the renormalization
channel we have introduced a numerical factor $g$ to account
for the fact that the projection of the general spectral density
will in general be different in the two channels. The
repulsive effective Coulomb interaction $\mus_{\ve{k},\ve{k}'}$
 is isotropic in an isotropic $s$-wave
formalism. The same holds for the `Hubbard' $U$ which is also
assumed to be large and isotropic. Thus, the effective
Coulomb potential is not expected to have a numerically
large $d$-wave symmetric part
and therefore does not contribute to the pairing channel in a
generalized $d$-wave formulation of the Eliashberg equations.

Within these
simplifying assumptions the Eliashberg equations
need first to be written on the imaginary
Matsubara frequency axis. They take on the following form \cite{Jiang}:
\begin{subequations}
\label{eq:ImagEli}
\bea
  \label{eq:ImagEliA}
  \tilde{\Delta}(i\omega_n;\theta) &=& g\pi T\sum\limits_m
  \cos(2\theta)\lambda(m-n)\nonumber\\
  &&\!\!\times
\left\langle\frac{\cos(2\theta')
  \tilde{\Delta}(i\omega_m;\theta')}{\sqrt{
  \tilde{\omega}^2(i\omega_m)+\tilde{\Delta}^2(i\omega_m;
  \theta')}}\right\rangle',\nonumber\\
\eea
for the renormalized pairing potential $\tilde{\Delta}(i\omega_n;\theta)$,
and
\bea
  \label{eq:ImagEliB}
  \tilde{\omega}(i\omega_n) &=& \omega_n+\pi T\sum\limits_m
  \lambda(m-n)\nonumber\\
  && \times
\left\langle\frac{\tilde{\omega}(i\omega_m)}
  {\sqrt{
  \tilde{\omega}^2(i\omega_m)+\tilde{\Delta}^2(i\omega_m;
  \theta')}}\right\rangle',\nonumber\\
 \eea
for the renormalized frequencies $\tilde{\omega}(i\omega_n)$.
Here, $\langle\cdots\rangle$
denotes the angular average over $\theta$. The quantity
$\lambda(m-n)$ has the usual form
\begin{equation}
  \label{eq:ImagEliC}
  \lambda(m-n) = 2\int\limits_0^\infty\,d\Omega\,
  \frac{\Omega I^2\chi(\omega)}{\Omega^2+(\omega_m-\omega_n)^2}.
\end{equation}
\end{subequations}
As written, Eqs. (\ref{eq:ImagEli}) do not depend on impurity
scattering. To include this possibility, we need to add into the
right hand side of Eq. (\ref{eq:ImagEliB}) a term of the form
\begin{equation}
  \label{eq:ImagEliD}
  \pi\Gamma^+\frac{\Omega(i\omega_n)}
  {c^2+\Omega^2(i\omega_n)+D^2(i\omega_n)}
\end{equation}
where $\Gamma^+$ is proportional to the impurity concentration
and $c$ is related to the electron phase shift for scattering off
the impurity. For unitary scattering, $c$ is equal to zero while
$c\to\infty$ gives the Born approximation, i.e.: the weak scattering
limit. In this case the entire impurity term reduces to the
form $\pi t^+\Omega(i\omega_n)$ with $c$ absorbed into $t^+$.
 To complete the specification
of Eq. (\ref{eq:ImagEliD}), we have
\begin{subequations}
\label{eq:ImagEliE}
\begin{equation}
  \label{eq:ImagEliE1}
  D(i\omega_n) = \left\langle\frac{\tilde{\Delta}(i\omega_n;\theta)}
  {\sqrt{
  \tilde{\omega}^2(i\omega_m)+\tilde{\Delta}^2(i\omega_m;
  \theta)}}\right\rangle, 
\end{equation}
and
\begin{equation}
  \label{eq:ImagEli2}
  \Omega(i\omega_n) = \left\langle\frac{\tilde{\omega}(i\omega_n)}
  {\sqrt{
  \tilde{\omega}^2(i\omega_m)+\tilde{\Delta}^2(i\omega_m;
  \theta)}}\right\rangle.
\end{equation}
\end{subequations}

While certain quantities, such as the penetration depth, can be
obtained quite directly from the numerical solution on the
imaginary frequency axis, i.e.: from $\tilde{\Delta}(i\omega_n;
\theta)$ and $\tilde{\omega}(i\omega_n)$, real frequency axis
solutions are needed to calculate the optical
conductivity. These equations for $\tilde{\Delta}(\nu+i\delta;
\theta)$ and $\tilde{\omega}(\nu+i\delta)$ with $\delta$
infinitesimal are more complicated and can
be written in the form \cite{schach1,schach2,schach3}:
\begin{widetext}
\begin{subequations}
\label{eq:57}
\begin{eqnarray}
  \tilde{\Delta}(\nu+i\delta;\theta) &=& \pi Tg
  \sum\limits_{m=0}^\infty\cos(2\theta)\left[\lambda(\nu-i\omega_m)+
  \lambda(\nu+i\omega_m)\right]\nonumber\\
 &&\times\left\langle
 {\tilde{\Delta}(i\omega_m;\theta')\cos(2\theta')\over
  \sqrt{\tilde{\omega}^2(i\omega_m)+\tilde{\Delta}^2(i\omega_m;
  \theta')}}\right\rangle'\nonumber\\
 &&+i\pi g\int\limits^\infty_{-\infty}\!dz\,\cos(2\theta)
  I^2\chi(z)\left[n(z)+f(z-\nu)\right]\times\nonumber\\
 &&\times\left\langle
  {\tilde{\Delta}(\nu-z+i\delta;\theta')\cos(2\theta')\over
  \sqrt{\tilde{\omega}^2(\nu-z+i\delta)-\tilde{\Delta}^2(\nu-z+i\delta;
  \theta')}}\right\rangle',
  \label{eq:57a}
\end{eqnarray}
for the pairing channel and
\begin{eqnarray}
  \tilde{\omega}(\nu+i\delta) &=& \nu+i\pi T%
  \sum\limits_{m=0}^\infty\left[\lambda(\nu-i\omega_m)-
  \lambda(\nu+i\omega_m)\right]\nonumber\\
  &&\times\left\langle
    {\tilde{\omega}(i\omega_m)\over
  \sqrt{\tilde{\omega}^2(i\omega_m)+\tilde{\Delta}^2(i\omega_m;
  \theta')}}\right\rangle'+\nonumber\\
  &&+i\pi\int\limits^\infty_{-\infty}\!dz\,
   I^2\chi(z)\left[n(z)+f(z-\nu)\right]\nonumber\\
  &&\times\left\langle
  {\tilde{\omega}(\nu-z+i\delta)\over
  \sqrt{\tilde{\omega}^2(\nu-z+i\delta)-\tilde{\Delta}^2(\nu-z+i\delta;
  \theta')}}\right\rangle'\nonumber\\
  && + i\pi\Gamma^+{\Omega(\nu)\over
   c^2+D^2(\nu)+\Omega^2(\nu)}
  \label{eq:57b}.
\end{eqnarray}
\end{subequations}
\end{widetext}
for the renormalization channel. Thermal factors appear
in these equations through the Bose and
Fermi distribution $n(z)$  and $f(z)$, respectively. Furthermore,
the abbreviations:
\begin{subequations}
\label{eq:59}
\begin{equation}
  \label{eq:59a}
  \lambda(\nu) = \int\limits^\infty_{-\infty}\!d\Omega\,
   {I^2\chi(\Omega)\over\nu-\Omega+i0^+},
\end{equation}
\begin{equation}
  \label{eq:59b}
  D(\nu) =  \left\langle{\tilde{\Delta}(\nu+i\delta;\theta)\over
  \sqrt{\tilde{\omega}^2(\nu+i\delta)-\tilde{\Delta}^2(\nu+i\delta;
  \theta)}}\right\rangle,
\end{equation}
\begin{equation}
  \label{eq:59c}
  \Omega(\nu) = \left\langle{\tilde{\omega}(\nu+i\delta)\over
  \sqrt{\tilde{\omega}^2(\nu+i\delta)-\tilde{\Delta}^2(\nu+i\delta;
  \theta)}}\right\rangle.
\end{equation}
\end{subequations}
have been used. Eqs. (\ref{eq:57}) are a set of nonlinear
coupled equations for the renormalized pairing potential $\tilde{\Delta}%
(\nu+i\delta;\theta)$ and the renormalized frequencies
$\tilde{\omega}(\nu+i\delta)$ with the superconducting gap given by
\begin{equation}
  \label{eq:60}
  \Delta(\nu+i\delta;\theta) = \nu\,{\tilde{\Delta}(\nu+i\delta;\theta)%
  \over\tilde{\omega}(\nu+i\delta)},
\end{equation}
or, if the renormalization function $Z(\nu)$ is introduced in the
usual way as $\tilde{\omega}(\nu+i\delta) = \nu Z(\nu)$, one finds
for the superconducting gap
\begin{equation}
  \label{eq:61}
  \Delta(\nu+i\delta;\theta) = {\tilde{\Delta}(\nu+i\delta;\theta)%
  \over Z(\nu)}.  
\end{equation}
These
equations are a minimum set and go beyond a BCS approach and
include the inelastic scattering known to be strong in the
cuprate superconductors.

From solutions of the generalized Eliashberg equations
we can construct the Green's function in Eq. (\ref{eq:para})
analytically continued to the real frequency axis $\Omega$.
In this formulation the expression for the in-plane
conductivity $\sigma_{ab}(T,\Omega)$
involves further averaging
over angles which needs to be done numerically. We find the following
result after further manipulations and rearrangements:
\begin{subequations}
\label{eq:62}
\begin{widetext}
\begin{eqnarray}
\lefteqn{\sigma_{ab}(\Omega) = {i\over\Omega}\frac{e^2N(0)v^2_F}
  {2}}\nonumber\\
 && \times\left\langle\int\limits_0^\infty\!{\rm d}\nu\,
  {\rm tanh}\left({\nu\over 2T}\right)
  \frac{1-N(\nu;\theta)N(\nu+\Omega;\theta)-
  P(\nu;\theta)P(\nu+\Omega;\theta)}
  {E(\nu;\theta)+E(\nu+\Omega;\theta)}\right.
  \nonumber\\
 &&\left.+\int\limits_0^\infty\!{\rm d}\nu\,
  {\rm tanh}\left({\nu+\Omega\over 2T}\right)
  \frac{1-N^\star(\nu;\theta) N^\star(\nu+\Omega;\theta)
  -P^\star(\nu;\theta) P^\star(\nu+\Omega;\theta)}
  {E^\star(\nu;\theta)+E^\star(\nu+\Omega;\theta)}\rd\nnp
 &&\ld +\int\limits_0^\infty\!{\rm d}\nu\,\left[{\rm tanh}
  \left({\nu+\Omega\over 2T}\right)-{\rm tanh}\left({
  \nu\over 2T}\right)\right]\right.\nonumber\\
 &&\left.\times
  \frac{1+N^\star(\nu;\theta) N(\nu+\Omega;\theta)
  +P^\star(\nu;\theta) P(\nu+\Omega;\theta)}
  {E(\nu+\Omega;\theta)-E^\star(\nu;\theta)}
  \right.\nonumber\\
 &&\left.+\int\limits_{-\Omega}^0\!{\rm d}\nu\,
  {\rm tanh}\left({\nu+\Omega\over 2T}\right)\left\{
  \frac{1-N^\star(\nu;\theta) N^\star(\nu+\Omega;\theta)
   -P^\star(\nu;\theta) P^\star(\nu+\Omega;\theta)}
  {E^\star(\nu;\theta)+E^\star(\nu+\Omega;\theta)}
  \right.\right.\nonumber\\
 &&\left.\left.
   +
  \frac{1+N^\star(\nu;\theta) N(\nu+\Omega;\theta)
  +P^\star(\nu;\theta) P(\nu+\Omega;\theta)}
  {E(\nu+\Omega;\theta)-E^\star(\nu;\theta)}
  \right\}\right\rangle,  \label{eq:62a}
\end{eqnarray}
\end{widetext}
with
\begin{equation}
  E(\omega;\theta) = \sqrt{\tilde{\omega}^2_{\bf k}
   (\omega)-\tilde{\Delta}^2_{\bf k}
   (\omega)}, \label{eq:62b}
\end{equation}
and
\bea
  N(\omega;\theta) &=& {\tilde{\omega}_{\bf k}(\omega)\over
   E(\omega;\theta)}, \label{eq:62c}\\
  P(\omega;\theta) &=& {\tilde{\Delta}_{\bf k}(\omega)\over
   E(\omega;\theta)}. \label{eq:62d}
\eea
\end{subequations}
In the above, the star refers to the complex conjugate.
This set of equations is valid for
the real and imaginary part of the conductivity as a function
of frequency $\Omega$. It contains only the paramagnetic
contribution to the conductivity but this is fine since we
have found that the diamagnetic contribution is small in
the case considered here.

The out-of-plane conductivity $\sigma_c(T,\Omega)$ at temperature
$T$ and frequency $\Omega$ is related to the current-current
correlation function $\Pi_c(T,i\nu_n)$ at the boson Matsubara
frequency $i\nu_n$
analytically continued to real frequency $\Omega$,
 and to the $c$-axis kinetic
energy $\langle H_c\rangle$ \cite{kim,radtke} via
\bea
  \label{eq:1}
  \sigma_c(T,\Omega) &=& {1\over\Omega}\left[
   \Pi_c(T,i\nu_n\to\Omega+i0^+)\right.\nonumber\\
   &&\left.-e^2d^2\langle H_c\rangle
   \right],
\eea
with $d$ the distance between planes in $c$-direction.
In terms of the in-plane thermodynamic Green's function
$\hat{G}({\bf k},i\omega_n)$ and for coherent hopping
$t_\perp({\bf k})$ perpendicular to the CuO$_2$ planes
\begin{widetext}
\begin{subequations}
  \label{eq:2}
  \begin{equation}
    \label{eq:2a}
   \Pi_c(T,i\nu_n) = 2(ed)^2T\sum\limits_{\omega_m}
   \sum\limits_{\bf k} t^2_\perp({\bf k})\,{\rm tr}\left\{
   \hat{\tau}_0\hat{G}({\bf k},i\omega_m)\hat{\tau}_0
   \hat{G}({\bf k},i\omega_m+i\nu_n)\right\}
  \end{equation}
and
\begin{equation}
  \label{eq:2b}
   \langle H_c\rangle = 2T\sum\limits_{\omega_m}
   \sum\limits_{\bf k} t^2_\perp({\bf k})\,{\rm tr}\left\{
   \hat{\tau}_3\hat{G}({\bf k},i\omega_m)\hat{\tau}_3
   \hat{G}({\bf k},i\omega_m)\right\}.
\end{equation}
\end{subequations}
\end{widetext}
In Eqs. (\ref{eq:2}) the $2\times2$ Nambu
Green's function $\hat{G}({\bf k},i\omega_m)$ describes
the in-plane  dynamics of the charge carriers with momentum
{\bf k} in the two dimensional CuO$_2$ plane Brillouin
zone and is given by
\begin{equation}
  \label{eq:3}
  \hat{G}({\bf k},i\omega_n) = {i\tilde{\omega}(i\omega_n)\hat{\tau}_0
  + \zeta_{\bf k}\hat{\tau}_3 +\tilde{\Delta}_{\bf k}(i\omega_n)
  \hat{\tau}_1\over -\tilde{\omega}^2(i\omega_n)-\zeta^2_{\bf k}-
  \tilde{\Delta}^2_{\bf k}(i\omega_n)},
\end{equation}
where the $\hat{\tau}$'s are Pauli $2\times2$ matrices,
$\zeta_{\bf k}$ is the band energy of the charge carriers
as a function of their momentum {\bf k}, $\tilde{\Delta}_{\bf k}%
(i\omega_n)$
is the renormalized pairing potential and $i\tilde{\omega}(i\omega_n)$
the renormalized Matsubara frequency. In our model these quantities are
determined as solutions of Eliashberg equations (\ref{eq:ImagEli}).

In Eqs.~(\ref{eq:2}) the out-of-plane matrix element $t_\perp({\bf k})$
can depend on the in-plane momentum {\bf k}. Models have been summarized
recently by Sandeman and Schofield \cite{sand} who
refer to previous literature \cite{xiang,anders,xiang1}. A possible
choice is $t_\perp({\bf k}) = t_\perp$, a constant. But, consideration
of the chemistry of the CuO$_2$ plane and of the overlap of one plane
with the next, suggests a form $t_\perp({\bf k}) = %
\cos^2(2\theta)$ where $\theta$ is the angle of {\bf k} in the
two dimensional CuO$_2$ Brillouin zone for the plane motion. This
matrix element eliminates entirely contributions from nodal quasiparticles
to the $c$-axis motion.

For incoherent impurity induced $c$-axis charge transfer
Eqs. (\ref{eq:2}) are to be modified. After an impurity
configuration average one obtains
\begin{widetext}
\begin{subequations}
\label{eq:4}
\begin{eqnarray}
  \Pi_c(T,i\nu_n) &=& 2(ed)^2T\sum\limits_m
   \sum\limits_{{\bf k},{\bf k}'}\overline{V^2_{{\bf k},{\bf k}'}}
   {\rm tr}\left\{
   \hat{\tau}_0\hat{G}({\bf k},i\omega_m)\hat{\tau}_0
   \hat{G}({\bf k}',i\omega_m+i\nu_n)\right\}\nonumber\\
   \label{eq:4a}\\
  \langle H_c\rangle &=& 2T\sum\limits_m
   \sum\limits_{{\bf k},{\bf k}'} \overline{V^2_{{\bf k},{\bf k}'}}
   {\rm tr}\left\{
   \hat{\tau}_3\hat{G}({\bf k},i\omega_m)\hat{\tau}_3
   \hat{G}({\bf k}',i\omega_m)\right\}\label{eq:4b},
\end{eqnarray}
\end{subequations}
\end{widetext}
with $\overline{V^2_{{\bf k},{\bf k}'}}$ the average of the square
of the impurity potential.
 If the impurity potential was taken to conserve momentum,
which it does not, we would recover Eqs. (\ref{eq:2}). Various
models could be taken for $\overline{V^2_{{\bf k},{\bf k}'}}$.
Here we use a form introduced by Kim \cite{kim} and
Hirschfeld {\it et al.} \cite{hirschf}
\begin{equation}
  \label{eq:4c}
  \overline{V^2_{{\bf k},{\bf k}'}} = \vert V_0\vert^2+
   \vert V_1\vert^2\cos(2\theta)\cos(2\theta'),
\end{equation}
with $\theta$ and $\theta'$ the directions of {\bf k} and
${\bf k}'$ respectively.

After analytic continuation to the real $\Omega$-axis
the real part of the incoherent conductivity along the $c$-axis is
given by (normalized to its normal state value
$\sigma_{1cn}$) \cite{hirschf}:
\bea
{\sigma_{1c}(\Omega)\over\sigma_{1cn}} &=&
  {1\over\nu}\int\!d\omega\,\left[f(\omega)-f(\omega+\Omega)
  \right]\nonumber\\
 &&\times\left[N(\omega+\Omega)N(\omega)\right.\nonumber\\
 &&\left.+\left\vert{V_1\over V_0}\right\vert
   P(\omega+\Omega)P(\omega)\right].\label{eq:inc}
\eea

\section{Application to High-$T_c$ Cuprates}
\subsection{The Compound YBa$_2$Cu$_3$O$_{6+\delta}$}
\subsubsection{The Normal State Infrared Conductivity}

We begin with a discussion of the normal-state scattering
of the charge carriers off spin fluctuations. For the spin
fluctuation spectral density $\itof{}$ we take a very simple
model motivated in the work of Millis {\it et al.} \cite{Pines1}
(MMP). We define a single characteristic  spin fluctuation
frequency $\omega_{SF}$ and take
\be
  I^2\chi(\omega) = I^2{\omega/\omega_{SF}\over\omega^2+
  \omega_{SF}^2},
  \label{eq:63}
\ee
where $I^2$ is a coupling constant which can be fit to normal-%
state infrared data based on Eq.~(\ref{eq:42}) with
$I^2\chi(\omega)$ playing the role of $\alpha^2F(\omega)$ in
Eq.~(\ref{eq:44}) for the self energy.

\begin{figure}[t]
\vspace*{-12mm}
\includegraphics[width=8cm]{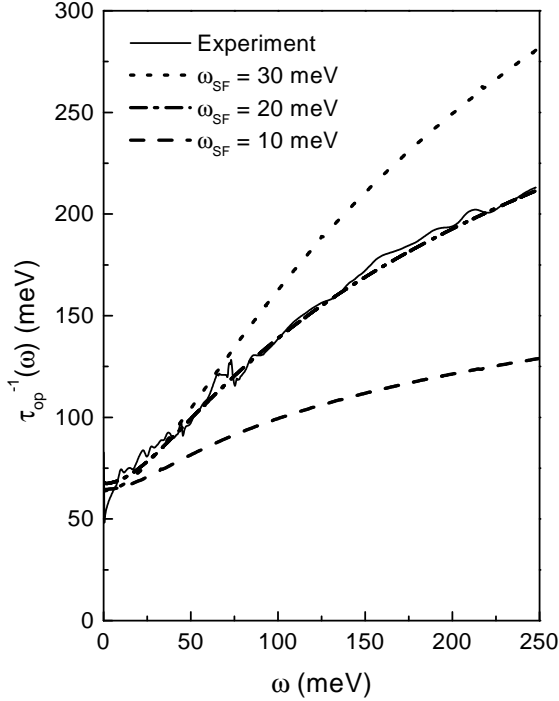}
\vspace*{-12mm}
\caption{
The normal-state optical scattering rate $\tau_{op}^{-1}(\omega)$
vs $\omega$ for YBCO with a $T_c = 92.4\,$K obtained from the
work of Basov {\it et al.} \protect\cite{Basov1} (solid line)
at a temperature of $95\,$K. The dash-dotted
curve from theory based on Eq.~(\protect\ref{eq:42}) with
an MMP model spectral density using a spin fluctuation frequency
$\omega_{SF} = 20\,$meV gives good agreement while the other choices
of $30\,$meV (dotted line) or $10\,$meV (dashed line) do not.
\label{fig:2}}
\end{figure}

\begin{figure}
\vspace*{-8mm}
\includegraphics[width=8cm]{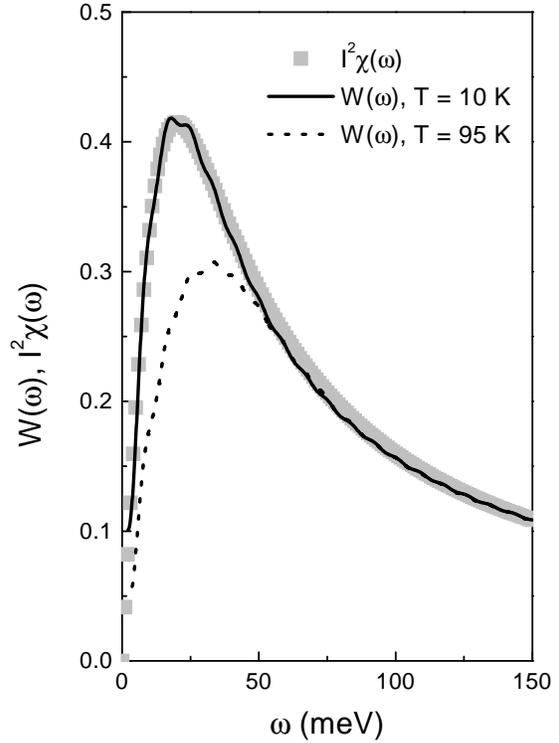}
\vspace*{-10mm}
\caption{
Comparison of the spectral density $\itof{}$ (gray solid squares)
in the MMP model \protect\cite{Pines1} ($\omega_{SF}=20\,$meV
with the function $W(\omega)$ defined
in Eq.~(\protect\ref{eq:47}) from the normal-state conductivity scattering
rate Eq.~(\protect\ref{eq:46}). The solid curve is at $T=10\,$K while
the dotted curve is at $95\,$K.
\label{fig:3}}
\end{figure}

In Fig.~\ref{fig:2} we show our result
for $\tau_{op}^{-1}(\omega)$ related to the
conductivity by Eq.~(\ref{eq:46}) for an optimally doped, twinned
YBa$_2$Cu$_3$O$_{6.95}$ (YBCO) single crystal
with $T_c=92.4\,$K at temperature $T=95\,$K. The solid
curve is the data of Basov {\it et al.} \cite{Basov1}. The
dashed and dotted curves are our best fits for
$\omega_{SF} = 10\,$meV and $30\,$meV, respectively, with
$I^2$ adjusted to get the correct absolute value of the
scattering rate at $T=95\,$K and low energies $\omega$.
We see that both values of
$\omega_{SF}$ do not give equally satisfactory fits to the data.
The dash-dotted curve, however, fits the data well and
corresponds to $\omega_{SF} = 20\,$meV. This fit provides
us with a model $\itof{}$ valid for the normal-state of YBCO.
A plot of this function is shown in Fig.~\ref{fig:3} as the
gray solid squares. Also shown in this figure are two sets
of theoretical results based on the $\itof{}$ with $\omega_{SF}=20\,$%
meV which serve to illustrate the inversion technique. The
experimental data
on $\tau_{op}^{-1}(\omega)$ gives the model $\itof{}$
spectrum. Next this
model spectrum is used in the normal-state conductivity
Eq.~(\ref{eq:42}) and two temperatures are considered, namely
$T=95\,$K (dotted curve) and $T=10\,$K (solid curve). Except for
some numerical noise it is clear that when $W(\omega)$
(solid curve) is computed
from Eq.~(\ref{eq:47}) which involves a second derivative of our
calculated data for $\tau_{op}^{-1}(\omega)$ computed from the
conductivity according to Eq.~(\ref{eq:46}), that the low temperature
data ($T=10\,$K) gives a remarkable accurate picture of the
input $\itof{}$ function and that to a very good approximation
$W(\omega)$ is the same as $\itof{}$ in this case. If the
temperature is increased to $95\,$K the match between $W(\omega)$
(dotted curve)
and $\itof{}$ is not quite as good. This shows that our
inversion technique summarized in Eqs.~(\ref{eq:46}) and
(\ref{eq:47}) is best when applied at low temperatures. The
resultant inverted $\itof{}$ gets smeared somewhat if high
temperatures are used instead.

\subsubsection{The Superconducting State, $\itof{}$ and the
Infrared Conductivity}

\begin{figure}
\vspace*{4mm}
\includegraphics[width=8cm]{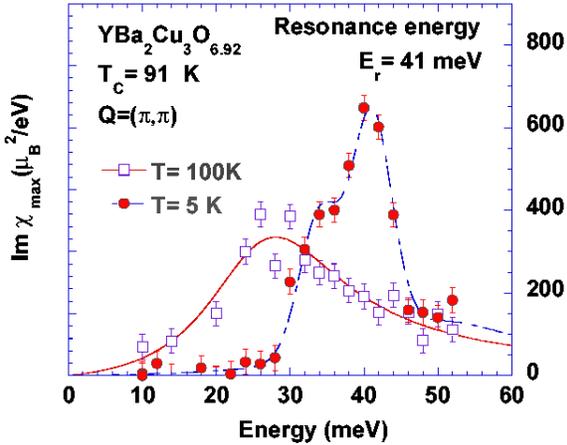}
\vspace{-6pt}
\caption{
Experimental results of the magnetic susceptibility at
$\ve{Q}=(\pi,\pi)$ in a sample of YBa$_2$Cu$_3$O$_{6.92}$.
The imaginary part of $\chi$ is measured as a function of
energy. Open squares are results at $T=100\,$K and
solid circles are at $T=5\,$K (in the superconducting state,
$T_c = 91\,$K for this sample). The energy of the spin
resonance $E_r = 41\,$ meV. Adapted from \protect\cite{Bourges}.}
\label{fig:4}
\end{figure}

Going to the superconducting state requires the solution of the
Eliashberg equations (\ref{eq:57}) and
evaluation of formula (\ref{eq:62}) for the optical
conductivity. This is much more complicated than the
corresponding normal-state analysis. Also, it is critical
to understand that since we are dealing with a highly
correlated system and the excitations we are exchanging in our
spectral density are within the electronic system itself, they
could be profoundly modified by the onset of the transition.
In Fig.~\ref{fig:4} we reproduce spin polarized inelastic
neutron scattering results of the spin fluctuations measured
in a sample of YBa$_2$Cu$_3$O$_{6.92}$ by Bourges {\it et al.}
\cite{Bourges} at two temperatures namely $T=100\,$K (open
squares) and $T=5\,$K (solid circles) which show the formation
of the $41\,$meV spin resonance in the superconducting
state. At $100\,$K the measured spectrum looks
very much like the simple spectrum used in our analysis of the
optical data in YBCO (Fig.~\ref{fig:3}) but this simple form
is profoundly modified in the superconducting state with the
imaginary part of the magnetic susceptibility depressed at small
$\omega$ and the formation of a sharp peak around $E_r=41\,$meV.
The possibility of a change in $\itof{}$ brought about by the
onset of superconductivity must be included in our analysis
of the optical data in the superconducting state. A question
we can immediately ask is: is the $41\,$meV peak seen in
optimally doped YBCO in neutron scattering also
seen in the superconducting state optical
conductivity? To perform the necessary analysis several modifications
of what has been done so far need to be considered. The observation
that, in the normal-state the well defined function $W(\omega)$
given in Eq.~(\ref{eq:47}) which is easily accessible when the
conductivity $\sigma(\omega)$ is known, can be identified to
a good approximation with $\itof{}$ may not hold in the
superconducting state of a $d$-wave superconductor.

\begin{figure}
\vspace*{-4mm}
\includegraphics[width=8cm]{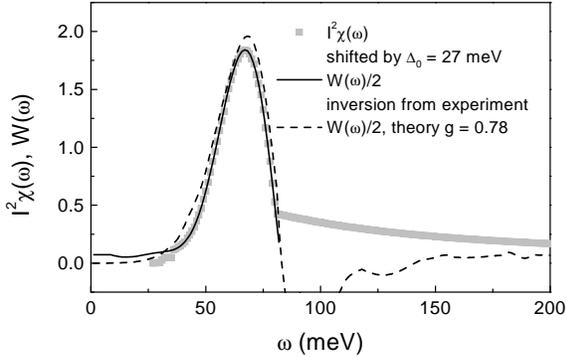}
\vspace*{-12pt}
\caption{
The model charge carrier spin excitation spectral density
$\itof{}$ for $T=10\,$K (gray solid squares) constructed from conductivity
data for optimally doped YBCO. The dashed line which follows the
gray solid squares faithfully, except for negative oscillations
just beyond the spin resonance around $68\,$meV, is
$W(\omega)/2$ obtained from our model $\itof{}$ (displaced
by the gap energy $\Delta_0 = 27\,$meV in the figure). The
solid line is the coupling to the resonance found directly
from experiment.
\label{fig:5}}
\end{figure}

It turns out from extensive calculations of the superconducting
state conductivity based on Eqs.~(\ref{eq:62}) by
Schachinger and Carbotte \cite{schach5,schach6} that a simple
modification of the rule
\[
  \itof{} \simeq W(\omega)
\]
can be found which applies approximately in the superconducting
state in the resonance region, namely
\[
  I^2\chi(\omega_1) \simeq {W(\omega_2)\over 2},
\]
with $\omega_2$ shifted by $\Delta_0(T)$, the energy gap at
temperature $T$, when compared with $\omega_1$.
This rule, while not exact, is nevertheless sufficiently accurate
to make it useful in obtaining first qualitative
information on the magnitude
and frequency dependence of the underlying charge carrier-exchange boson
interaction spectral density from optical data. We expect the
changes in going to the superconducting state to be the growth
of the resonance, while at higher energies there should be no change in
$\itof{}$ from its normal state value. A detailed comparison
of $\itof{}$ and $W(\omega)$ in the superconducting state
will be given later when we relate
additional structures in $W(\omega)$ not in $\itof{}$ to structures
in the superconducting quasiparticle density of states which
introduce distortions in $W(\omega)$ as compared to the
underlying spectral density. In Fig.~\ref{fig:5} we
show results for the case of YBCO \cite{schach4,puchkov} at
$T = 10\,$K. The solid line was obtained directly from differentiation of
experimental data on the optical scattering rate using the
superconducting state conductivity $\sigma(\omega)$ in
Eq.~(\ref{eq:46}) and the definition (\ref{eq:47}) of
$W(\omega)$. It is limited to the resonance region.
It shows that $W(\omega)/2$ clearly
has a peak corresponding to the peak seen in the spin susceptibility
of Fig.~\ref{fig:4} measured in neutron scattering. Of course,
$\itof{}$ includes the coupling constant $I^2$ between electrons
and spin fluctuations and is not strictly the imaginary part of the
spin susceptibility. These two functions are not the
same but we do know that an optical resonance peak does appear
when the neutron peak is observed. They fall at the same
frequency $E_r$ and look similar in other aspects.
 It is important to emphasize that this
solid curve comes directly from the data on $\sigma(\omega)$
and can be interpreted as evidence for
coupling of the charge carriers to the spin one resonance
at $41\,$meV. The other curves in the figure are equally important.
The gray squares represent the $\itof{}$ used in calculations
displaced in energy by the gap $\Delta_0 = 27\,$meV. It is
constructed completely from experiment. The fitting procedure
involves two critical independent steps. First the data on
$\tau_{op}^{-1}(\omega)$ in the normal-state is used to get a
background spectrum of the form given in Eq.~(\ref{eq:63}) which
applies to the normal-state. This defines $\omega_{SF}$ and the
corresponding $I^2$. This spectrum is also valid at the critical
temperature $T_c$. We use this to determine the last
parameter, the anisotropy parameter $g$ in Eqs.~(\ref{eq:ImagEliA})
and (\ref{eq:57a}),
in solving the linearized Eqs.~(\ref{eq:ImagEli}) for $g$ to give
the required $T_c$. In the second step the normal-state result
for the spectral density is modified only in the region of the
resonance peak leaving it unchanged at higher energies. The
resonance is positioned and its magnitude given by the data for
the experimental $W(\omega)/2$ (solid curve). There is no
ambiguity and the procedure is definite. A check on the consistency
of this procedure is then performed in calculating the
theoretical $W(\omega)/2$ (dashed line) from the theoretical
optical scattering rate calculated numerically from the solutions
of the Eliashberg equations. We see that the theoretical data
agree remarkably well with experiment in the region of the
resonance. More explanations of the differences between $W(\omega)/2$
and $\itof{}$ beyond the resonance region will be provided
later on.

\begin{figure}
\vspace*{-6mm}
\includegraphics[width=8cm]{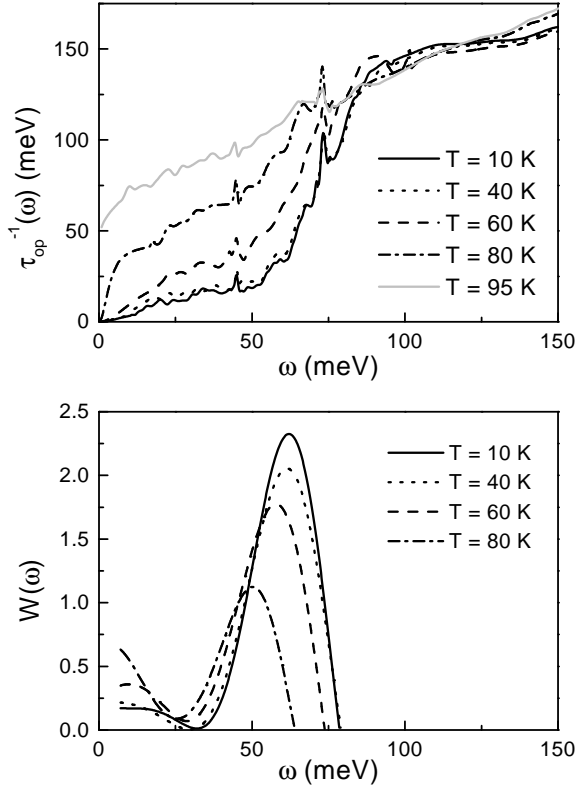}
\vspace*{-5mm}
\caption{
Top frame: optical scattering rate $\tau^{-1}_{op}(T,\omega)$ in
meV for optimally doped, twinned YBCO single crystals
\protect{\cite{Basov1}}. Bottom
frame: function $W(\omega)$ vs $\omega$ in the region of the
optical resonance.} 
\label{fig:6}
\end{figure}

After this important low temperature consistency check
we can now study
in more detail the temperature dependence of the
spectral density $\itof{}$ in the superconducting
state of an optimally doped, twinned
YBCO single crystal using experimental data. The frequency dependence of the
optical scattering rate has been studied at five temperatures
by Basov {\it et al.} \cite{Basov1}. We reproduce their experimental
results in the top frame of Fig.~\ref{fig:6}. The data are
for $T=10\,$K solid line, $T=40\,$K dotted line, $T=60\,$K
dashed line, $T=80\,$K dash-dotted line, and $T=95\,$K (in the
normal-state) gray solid line. We see that in the normal-state
$\tau^{-1}_{op}(\omega)$ vs $\omega$ shows a quasi-linear
dependence on $\omega$ but in all other curves a depression
below the normal curve is seen at small $\omega$ below roughly
$75\,$meV. The depression is the more pronounced the lower the
temperature. At higher frequencies all curves roughly coincide.
At low temperatures, $\tau_{op}^{-1}(\omega)$
as a function of $\omega$
shows a small value up to $50\,$meV or so, with a sharp rise
around $75\,$meV characteristic of the existence of a sharp
peak in $\itof{}$. It is clear that this peak increases in
strength as $T$ is lowered into the superconducting state.
More quantitative information on the temperature variation
of the optical resonance, its strength in $\itof{}$, and its
position is shown in the bottom frame of Fig.~\ref{fig:6}
where we show the second derivative $W(\omega)$ function derived
directly from the data given in the top frame by performing the
differentiation indicated in Eq.~(\ref{eq:47}).
The solid curve is for $T=10\,$K, the dotted curve for
$T=40\,$K, the dashed for $T=60\,$K, and dash-dotted for
$T=80\,$K. The height of the peak of the resonance clearly
increases with lowering of the temperature. In the curves for
$W(\omega)$ vs $\omega$ the position of the peak is seen to
be reduced as $T$ is increased towards $T_c$. If it is
remembered that the peak in $W(\omega)$ is located at the gap
value $\Delta_0(T)$ plus the position of the resonance $E_r$ and the
temperature dependence of the gap is accounted for (it decreases
with increasing temperature and rapidly goes to zero as
$T_c$ is approached) then we conclude that the position of the
resonance is temperature independent although its strength
decreases as $T$ increases. This is also in agreement
with the observation made by Dai {\it et al.} \cite{Dai}
that the energy at which the spin one resonance is observed
in YBCO ($41\,$meV) is temperature independent.

\begin{figure}
\vspace*{-2mm}
\begin{center}
\includegraphics[width=8.5cm]{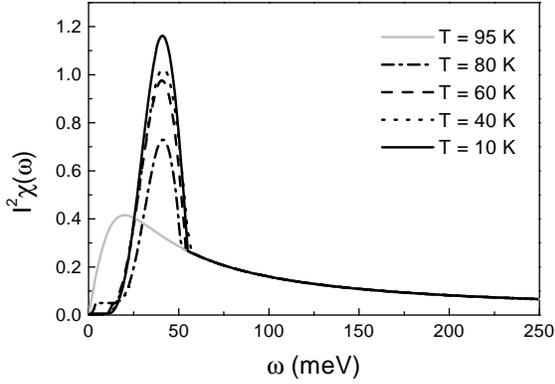}
\end{center}
\vspace*{-6mm}
\caption{
The charge carrier-spin excitation spectral density $\itof{}$
determined from optical scattering data at various temperatures.
Solid gray curve $T=95\,$K, dash-dotted $T=80\,$K, dashed
$T=60\,$K, dotted $T=40\,$K, and black solid $T=10\,$K. Note
the growth in strength of the $41\,$meV optical resonance as the
temperature is lowered.}
\label{fig:7}
\end{figure}

\begin{figure}
\vspace*{-5mm}
\includegraphics[width=8cm]{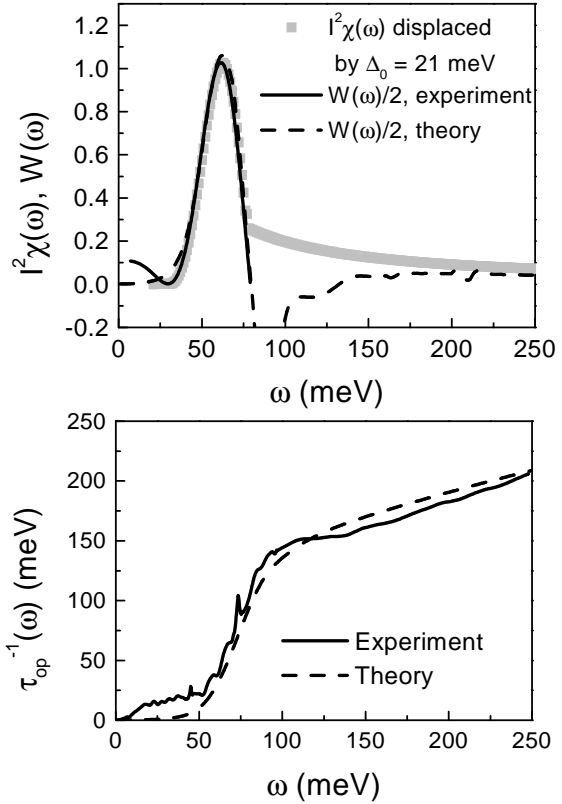}
\vspace*{-4mm}
\caption{
The top frame gives the resonance peak obtained directly from
the optical data for YBCO in the superconducting state (solid curve)
at $T=40\,$K,
the model $\itof{}$ (gray squares) displaced by the gap, and
the theoretical results for $W(\omega)/2$ obtained from the
model spectral density (dashed line). The bottom frame gives
the optical scattering rate $\tau^{-1}_{op}(\omega)$ vs
$\omega$. The experimental results give the solid curve and
our theoretical fit to it is the dashed curve.
}
\label{fig:8}
\end{figure}

In Fig.~\ref{fig:7} we show results for the spectral density
$\itof{}$ vs $\omega$ obtained from the data of Fig.~\ref{fig:6}
at five temperatures, namely $T=95\,$K (solid gray curve),
$T=80\,$K (dash-dotted curve), $T=60\,$K (dashed curve),
$T=40\,$K (dotted curve), and $T=10\,$K (solid black curve).
The procedure follows in all cases the procedure already
described in detail for the $T=10\,$K data. Now, the resonance
is positioned and its magnitude given by the $W(\omega)$ data
shown in the bottom frame of Fig.~\ref{fig:6} and $W(\omega)/2$
is used to modify the normal-state MMP background spectrum
only in the region of the resonance. A consistency check is
then performed for all temperatures and we present the
$T=40\,$K result in Fig.~\ref{fig:8}. The top frame gives
$W(\omega)/2$ obtained directly from experiment
in the resonance region (solid line,
corresponds to the dotted line in the bottom frame of
Fig.~\ref{fig:6}). The model $\itof{}$ for this temperature
is given by the gray solid squares displaced in energy by the
superconducting gap $\Delta_0=21\,$meV. This spectrum agrees
with the experimental $W(\omega)/2$ in the resonance region.
The dashed curve, finally, represents the theoretical
$W(\omega)/2$ and we recognize, again, a remarkable agreement
between theory and experiment in the appropriate energy range. The fit
to the optical scattering rate which in the end is the quantity
that matters is shown in the bottom frame of Fig.~\ref{fig:8}.
The theoretical curve
(dashed line) follows well the experimental data (solid line).
In this
sense we have found a spectral density which can reproduce
the measured optical scattering rate at $T=40\,$K and
it does not matter much how we arrived at our final model
for $\itof{}$ which involved, as a step, consideration of the
function $W(\omega)$ which served to guide our choice.

\begin{figure}
\vspace*{-27mm}
\includegraphics[width=8cm]{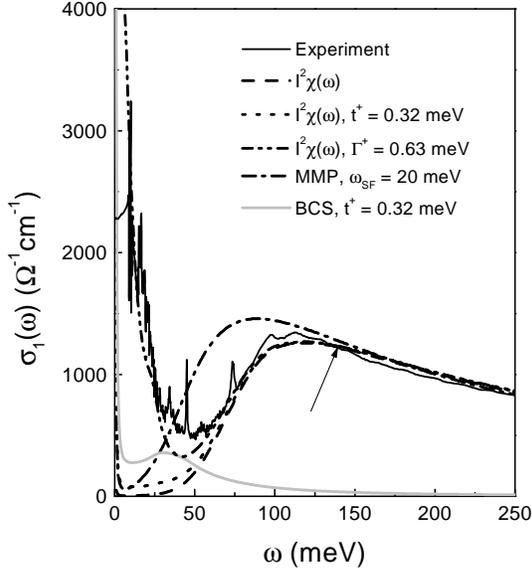}
\vspace*{-14mm}
\caption{
Comparison of the real part of the in-plane conductivity
$\sigma_1(\omega)$ vs $\omega$ for various models ($T=10\,$K). The grayed
solid line with a peak before $50\,$meV is BCS. The dash-dotted
line is an Eliashberg calculation
with an MMP spectral density peaked at $\omega_{\rm SF} = 20\,$meV.
The dashed line is the same but with our temperature and frequency
dependent $I^2\chi(\omega)$ (see Fig.~\protect\ref{fig:7}) used
instead of the MMP model. As described in the text this
charge carrier-exchange
boson interaction spectral density $I^2\chi(\omega)$ has been
determined through a consideration
of in-plane optical data. The dotted (Born) and dash-double-dotted
(unitary scattering) curves
include impurities in addition to the $I^2\chi(\omega)$ model for
inelastic scattering. The solid line
is the data of Homes {\it et al.} \protect\cite{homes}.
\label{fig:9}}
\end{figure}

We return now to a more detailed discussion of the optical conductivity.
In Fig.~\ref{fig:9} we show the real part of the optical conductivity
$\sigma_1(\omega)$ for
untwinned, optimally doped YBCO single crystals (solid line) reported
by Homes {\it et al.} \cite{homes} and compare with results of
various calculations. This means that we are now comparing
theoretical predictions with experimental YBCO data which
have not been used to derive the $\itof{}$ spectra.
The gray solid curve is the BCS result
for a gap of $24\,$meV and impurity scattering in Born approximation
corresponding to $t^+ = 0.32\,$meV. It is very clear that no
agreement with experiment is possible with BCS $d$-wave. One needs
to go to an Eliashberg formulation if one is to even get close
to the data. In some sense this is very positive since a good fit
with a BCS formulation would mean that details of the pairing
potential do not play an important role in the conductivity and,
thus, our procedure would not
be a good way to pin down some of the details
of the pairing interaction. The dash-dotted curve represents
results of Eliashberg calculations but with an MMP model for the
$\itof{}$ kernel. This ignores the growth of the $41\,$meV optical
resonance that enters when the superconducting state develops.
While the agreement with the data is good at high energies
beyond $100\,$meV say, it fails completely in the low energy
region. In particular, the Drude like peak at very low energies
is much too narrow. It is important, however, to emphasize the
difference between Eliashberg and BCS at high energies where
BCS gives a conductivity which is much too small while, in
comparison, Eliashberg with an MMP kernel gives larger values
reflecting the long tails extending to $400\,$meV in the MMP
$\itof{}$ spectrum. This is taken as strong evidence for the
existence of long tails in the pairing spectral density and
argues against a pure phonon mechanism which would be lot more
confined in energy.

The dashed curve in Fig.~\ref{fig:9} gives our results for the
real part of the conductivity $\sigma_1(\omega)$ when our
$\itof{}$ at $T=10\,$K (solid line in Fig.~\ref{fig:7})
is used in the calculations rather than
the MMP kernel. The existence of the $41\,$meV resonance shifts
the large rise in the conductivity which now begins at much higher
energies $\sim 50\,$meV than in the MMP case.
It also leads to a maximum around $100\,$meV in
good agreement with experiment. Even better agreement can be
obtained if a small amount of impurity scattering is
included within the unitary or resonant scattering limit
$c\to 0$ in Eq.~(\ref{eq:57}). Results with $\Gamma^+=0.63\,$%
meV are shown as the dash-double-dotted curve which displays all
the important characteristics observed in the experimental data.
The agreement is truly very good. A final curve including only
Born impurity scattering (dotted curve) shows that this limit
cannot explain the data. It should be clear from this comparison
that BCS theory is quite inadequate in describing the
observed features of the real part of the infrared conductivity
as a function of frequency while Eliashberg theory can give a
good fit. It is also clear that some impurity scattering in the
unitary limit is needed and that the electron-boson exchange
spectral density has long tails extending to at least
$400\,$meV, and at $T=10\,$K has a large contribution
from the $41\,$meV resonance peak.

\begin{figure}
\vspace*{2mm}
\includegraphics[width=8cm]{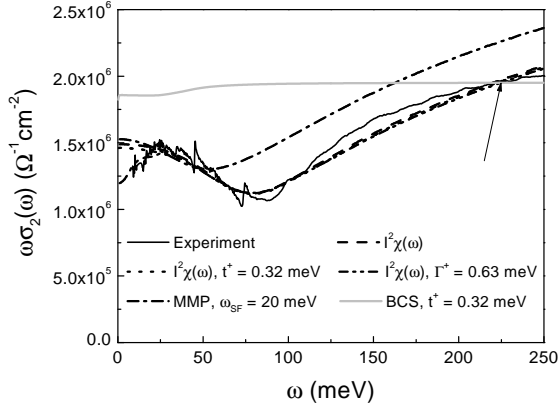}
\vspace*{-7mm}
\caption{
The imaginary part of the conductivity $\omega\sigma_2(\omega)$
vs $\omega$ for the various models described in Fig.~%
\protect\ref{fig:9} ($T=10\,$K). The
solid curve is the data \protect\cite{homes}.
The dash-dotted curve is the result of
an Eliashberg calculation with an MMP model while the other curves
are based on the model $I^2\chi(\omega)$ which includes the
$41\,$meV resonance. These three curves are for the pure case
(only inelastic scattering, dashed line), the others are
with some additional
elastic impurity scattering in Born (dotted) and unitary (dash-double-dotted)
limit with $t^+ = 0.32\,$meV and $\Gamma^+ = 0.63\,$meV respectively.
The gray solid line is the BCS result.}
\label{fig:10}
\end{figure}

In Fig.~\ref{fig:10} we show our results for the imaginary part
of the optical conductivity $\sigma_2(\omega)$.
What is plotted is $\omega\sigma_2(\omega)$
vs $\omega$ at $T=10\,$K. The solid curve gives data from
Homes {\it et al.} \cite{homes}. The dash-dotted curve are
Eliashberg results based on an MMP kernel which does not account for
charge carrier coupling to the $41\,$meV resonance. It fails to reproduce
the data. On the other hand, when the resonance is included in
our $\itof{}$ we get the dashed curve which agrees with the
data much better and shows a large depression in the curve located
around $75\,$meV in agreement with the data. This is the
signature in $\sigma_2(\omega)$ of the $41\,$meV resonance
in $\itof{}$. At very low
energies the agreement is not as good. As for the real part of the
conductivity this region is sensitive to impurity scattering
while at higher energies only inelastic scattering is really
important. The dash-double-dotted curve includes impurities in
the unitary limit with $\Gamma^+=0.63\,$meV as before. This
produces excellent agreement with the data even in the small
$\omega$ region which is not as well described
in the case of Born scattering
(dotted curve with $t^+=0.32\,$meV). It is clear from this graph
that $\omega\sigma_2(\omega)$ has an easily identifiably
signature of the spin resonance. Also, the spectral density
extends to high energies and Eliashberg theory with some
contribution from unitary scattering impurities is in
very good agreement with the data.

While we have seen that an unmistakable signature of the
$41\,$meV optical resonance exists in the data on both real and
imaginary part of the infrared conductivity as a function of
energy $\omega$ in the $T=10\,$K data, the resonance is even
clearer in the second derivative defined by $W(\omega)$ of
Eq.~(\ref{eq:47}). The near equality between $W(\omega)$ and
$\atof{}$ established by Marsiglio {\it et al.} \cite{mars4}
for a phonon mechanism was only for the normal-state, i.e.:
$\tau^{-1}_{op}(\omega)$ entering the formula is the normal-%
state optical scattering rate.  But in the high $T_c$ cuprates the
only low temperature data available is often in the
superconducting state. Thus, we need to discuss in more detail
what happens in Eq.~(\ref{eq:47}) when $\tau^{-1}_{op}(\omega)$
is replaced by its superconducting state value.
As we have seen in Fig.~\ref{fig:3}, in the normal-state and at
low temperatures $W(\omega)$ is almost exactly \cite{schach4,%
schach5,schach6,schach7,schach7a} equal to the input
$I^2\chi(\omega)$ for models based on the NAFFL.
Of course, $I^2\chi(\omega)$ is seen in
$W(\omega)$ through electronic processes. But in the normal-%
state the electronic density of states $N(\E)$ is
constant in the important energy region
and, thus, does not lead to additional structures in
$W(\omega)$ that are not in $I^2\chi(\omega)$.
Such additional structures would then corrupt
the signal, if the aim is to obtain $I^2\chi(\omega)$ from $W(\omega)$.
This is no longer the case in the superconducting
state because of the logarithmic van Hove singularities in
$N(\varepsilon)$ and these do indeed strongly influence the
shape of $W(\omega)$ and introduce additional structures in
$W(\omega)$ corresponding
to combinations of the positions of the singularity in
$N(\varepsilon)$ and the peak in $I^2\chi(\omega)$ at
$E_r$ as described
by Abanov {\it et al.} \cite{abanov} The structures in $W(\omega)$
corresponding to these singularities contaminate the signal
in the sense that $W(\omega)$ in the superconducting state is no longer
equal to the input $I^2\chi(\omega)$ \cite{schach5,schach6}.
In fact, only the resonance peak appears clearly at $\Delta_0+%
E_r$ and its size in $W(\omega)$ is about twice the value
of $I^2\chi(\omega)$ at that frequency. In some cases the
tails in $W(\omega)$ also match well the tails in $I^2\chi(\omega)$.
In the end, of course, $W(\omega)$ serves only as a guide and it
is the quality of the final fit to the conductivity data that
determines the quality of the derived $I^2\chi(\omega)$.

\begin{figure}
\vspace*{-15mm}
\includegraphics[width=8cm]{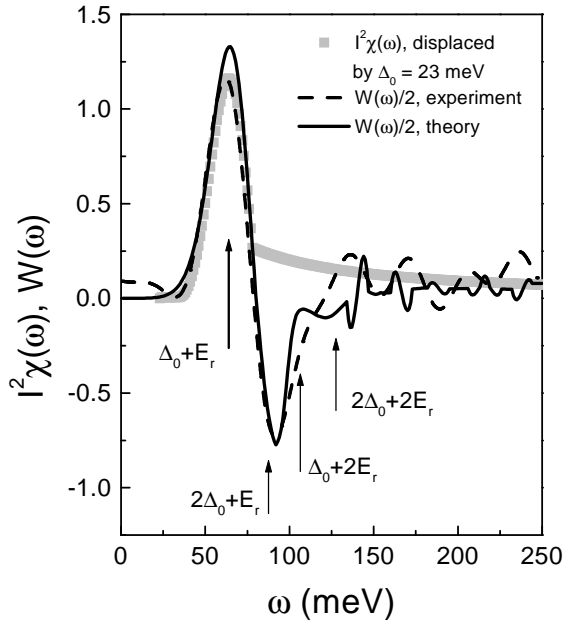}
\vspace*{-16mm}
\caption{
Second derivative $W(\omega)$ compared with the input
spectral density $I^2\chi(\omega)$. The $41\,$meV peak in $I^2\chi(\omega)$
(gray squares) is clearly seen in $W(\omega)/2$ (solid line) as are the
tails at higher energies. In the energy region between 75 and $150\,$%
meV the van Hove singularities in the electronic density of states
show up added on to $E_r$ and distort the correspondence
between $W(\omega)/2$ and $I^2\chi(\omega)$. The dashed line shows
the function $W(\omega)/2$ derived from experimental data.}
\label{fig:11}
\end{figure}

Nevertheless, besides giving a measure of the coupling of the
charge carriers to the optical resonance $W(\omega)$ can also be
used to see the position of density of states singularities, as shown in
Fig.~\ref{fig:11} where $I^2\chi(\omega)$ (gray squares) and
$W(\omega)$ (solid line) derived from our theoretical results
are compared. Also shown by vertical arrows are the positions
of $\Delta_0+E_r$, $2\Delta_0+E_r$,
$\Delta_0+2E_r$, and $2\Delta_0+2E_r$. We note
structures at each of these places  and this information is
valuable. Note that at $2\Delta_0+E_r$ the large negative
oscillation seen in $W(\omega)$ is mainly caused by the kink
in $I^2\chi(\omega)$ (gray squares) at about $75\,$meV.
 The density of electronic states effects clearly
distorted the spectrum above the resonance peak
and $W(\omega)$ stops agreeing with the input $I^2\chi(\omega)$
in this region until about $150\,$meV where agreement is
recovered. In summary, $W(\omega)$ contains some information
on singularities in $N(\varepsilon)$ as well as on the shape and size of
$I^2\chi(\omega)$ and, in the superconducting state, the two effects
cannot be clearly separated. Nevertheless, $W(\omega)$
remains a valuable intermediate step in the construction of
a charge carrier-exchange boson interaction spectral density from
optical data. To close, the dashed curve in Fig.~\ref{fig:11} is
the direct experimental data for $W(\omega)/2$ which is remarkably
similar to theory when we consider that a second derivative is
needed to get this curve.

\begin{figure}
\vspace*{3mm}
\includegraphics[width=8cm]{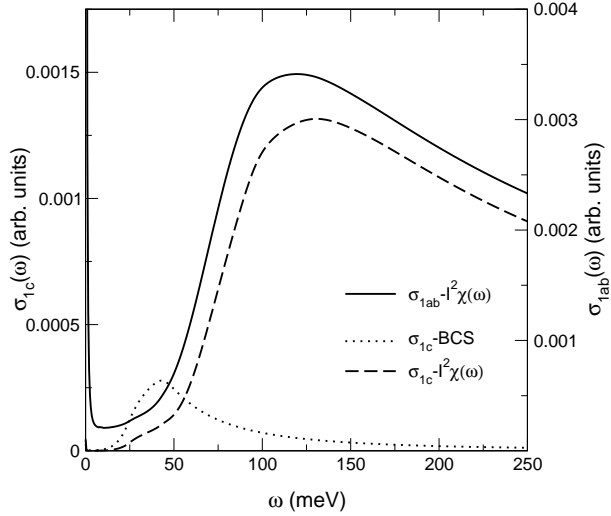}
\caption{Comparison between in-plane (solid line) and out
of plane (dashed line) real part of the $c$-axis conductivity
$\sigma_{1c}(\omega)$ vs $\omega$ in an Eliashberg model
with our model carrier-boson spectral
density $I^2\chi(\omega)$ which includes the $41\,$meV spin
resonance. The dotted curve is $\sigma_{1c}(\omega)$
for a BCS $d$-wave
model with the same gap value as in the Eliashberg
work and is included for comparison.}
\label{fig:12}
\end{figure}

Next we want to concentrate on the out-of-plane conductivity
($c$-axis conductivity) of YBCO. Before presenting results
we stress that the boson exchange kernel $I^2\chi(\omega)$ is an
in-plane quantity and is taken from our discussion of the in-plane
conductivity. It is not fitted to any $c$-axis data. It is
to be used unchanged to calculate the out-of-plane
conductivity assuming first coherent hopping with
$t_\perp({\bf k}) = t_\perp\cos^2(\phi)$ in Eqs.~(\ref{eq:2}).
The solid curve in Fig.~\ref{fig:12} is the
in-plane Eliashberg result which is included for comparison with the
dashed curve which is for the $c$-axis. In the boson assisted
region, which would not exist in a BCS theory, both curves
have a remarkably similar behavior. At very low frequencies,
a region which comes mainly from the coherent delta function part of
the carrier spectral density, and which is the only part included in
BCS, we note a narrow Drude-like peak in the solid curve.
This part is suppressed in the $c$-direction
(dashed curve) because the contribution from the nodal
quasiparticles are effectively left out by the
$t_\perp\cos^2(\phi)$ weighting term. Also, shown for comparison
are BCS results for coherent hopping (dotted
curve) along the $c$-axis. These results show no resemblance to our
Eliashberg results and also do not agree with
experiment Fig.~\ref{fig:13}. What determines the main rise in the region
beyond the Drude part of the conductivity in $\sigma_{1c}(\omega)$
are the boson assisted processes and this rise does not signal
the value of the gap or twice the gap for that matter, but rather
a combination of $\Delta_0$ and the resonance energy $E_r$.

In Fig.~\ref{fig:13} we compare the data of Homes {\it et
\begin{figure}
\vspace*{4mm}
\includegraphics[width=8cm]{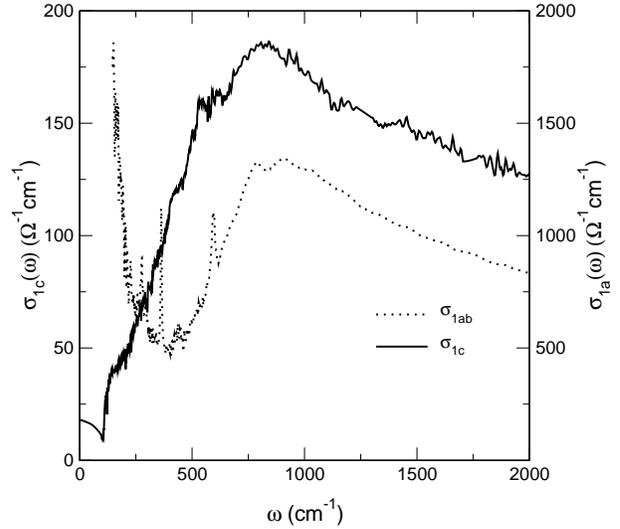}
\caption{Comparison between in-plane (dotted) and out
of plane (solid) for the real part of the conductivity $\sigma_1(\omega)$
vs $\omega$. The data is from Homes {\it et al.} \protect{\cite{homes}}.}
\label{fig:13}
\end{figure}
al.} \cite{homes} on the same graph for in-plane (dotted)
and out-of-plane (solid) conductivity $\sigma_1(\omega)$.
It is clear that in the $c$-direction, the nodal
quasiparticles seen in the dotted curve are strongly
suppressed. This favors the $t_\perp\cos^2(2\phi)$
matrix element for the $c$-axis dynamics as we have just seen.
Further, in the
boson assisted region the two curves show almost perfect
agreement with each other, which again favors the $t_\perp\cos^2(2\phi)$
coupling as was illustrated in the theoretical curves of
Fig.~\ref{fig:12}. One difference is that the main rise,
indicating the onset of the boson assisted incoherent
(in-plane) processes, appears to have shifted slightly
toward lower
frequencies in the $c$-axis data as opposed to a shift
to slightly higher frequencies in our theory. It should
be remembered, however, that in the raw $c$-axis data,
large structures appear in the conductivity due to
direct phonon absorption and these need to be subtracted
out, before data for the electronic background of Fig.~\ref{fig:13}
can be obtained. In view of this, it is not clear to us how
seriously we
should take the relatively small disagreements that we
have just described between theory and experiment.

\begin{figure}
\vspace*{-30mm}
\includegraphics[width=8cm]{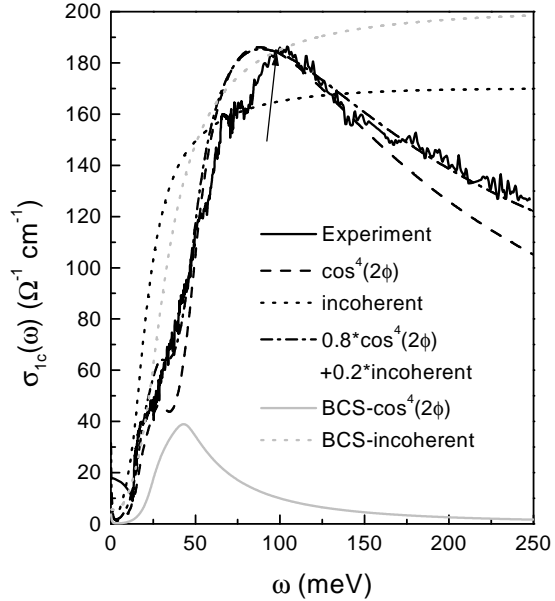}
\vspace*{-8mm}
\caption{Comparison with the data of Homes {\it et al.}
\protect{\cite{homes}} for the $c$-axis conductivity (black solid curve).
The theoretical curves were obtained in a BCS theory,
solid gray (coherent), dotted gray (incoherent) and the
others in Eliashberg theory with MMP model and impurities
$t^+ = 0.32\,$meV. The black dotted curve is for incoherent $c$-axis
with $\vert V_1/V_0\vert = 1$, the dashed for coherent
$c$-axis with $t_\perp(\phi) = t_\perp\cos^2(2\phi)$
with $\phi$ an angle in the two dimensional CuO$_2$
Brillouin zone, and the dash-dotted is a fit to the data
provided by a mixture of coherent and incoherent. We
stress that this last fit is for illustrative purposes only,
and is not unique.}
\label{fig:14}
\end{figure}

With the above
reservation kept in mind we show in Fig.~%
\ref{fig:14}, a comparison of various theoretical results with
experimental $c$-axis conductivity (black solid
line). There are five additional curves. The black ones
are obtained from an Eliashberg calculation based on the
MMP model for $I^2\chi(\omega)$ with impurities $t^+ =%
0.32\,$meV included to simulate the fact that the samples
used are not perfect, i.e.: are not completely pure, but this
parameter does not play a critical role in this discussion.
Incoherent $c$-axis coupling is assumed with
$\vert V_1/V_0\vert = 1$ (black dotted). It is clear that this
curve does not agree well with the data and that the coupling
along the $c$-axis cannot be dominated by incoherent hopping
between planes.
 This is also in agreement with the results
of a theoretical study by Dahm {\it et al.} \cite{dahm}
who also observed better agreement for coherent $c$-axis
conductivity in the overdoped regime.
On the other hand the fit with the black dashed line is good in comparison.
It uses the same MMP model but with
coherent coupling of the form $t_\perp({\bf k}) = t_\perp\cos^2(2\phi)$.
This fit may already be judged satisfactorily but it
should be remembered that if we had used the model of
$I^2\chi(\omega)$ with the $41\,$meV peak included instead
of MMP, the agreement would have deteriorated.
This is troubling since one
would expect that coupling to the $41\,$meV spin
resonance would be stronger in the $c$-direction data than
it is in the in-plane data. This is because the $c$-axis
emphasizes the hot spots around the antinodal directions
which connect best to $(\pi,\pi)$ in the magnetic susceptibility.
This is the position in momentum space where this spin
resonance is seen to be located in optimally doped YBCO.
On the other hand, recent ARPES data \cite{Varma,abrah1,valla,%
kamen} which fit well the MFL (marginal Fermi liquid)
phenomenology show little
in-plane anisotropy for scattering around the Fermi surface
and this is consistent with the findings here.

The dash-dotted curve in Fig.~%
\ref{fig:14} illustrates a fit to the data that can be achieved with a
dominant coherent piece and subdominant incoherent
contribution. It is not clear to us whether such a close fit
is significant given the uncertainties in the data and
the lack of uniqueness in the fitting procedure. It does, however,
illustrate the fact that a small
amount of incoherent $c$-axis hopping cannot be completely
ruled out from consideration of the infrared data and that
this data can be understood quite well within Eliashberg
theory.
The last two curves (solid gray and dotted gray) are based on
BCS $d$-wave theory and are reproduced here to illustrate the
fact that such a theory is unable to explain the $c$-axis
data. The solid gray curve is with $t_\perp({\bf k}) =
t_\perp\cos^2(2\phi)$ and the dotted gray one for incoherent
$c$-axis transport. Compared with our Eliashberg results the
agreement with the data is poor.

\subsubsection{The Microwave Conductivity}

The microwave conductivity as a function of temperature in
pure single crystals of YBCO revealed the existence of a very
large peak around $40\,$K \cite{BonnA} whose size and position
in temperature depends somewhat on the microwave frequency
used. This peak has been widely interpreted as due to a
rapid reduction in the inelastic scattering below $T_c$ and
is generally referred to as the collapse of the low-temperature
inelastic scattering rate.
This has been taken as strong evidence that the mechanism
involved is electronic in origin and,
this fact translates in our formalism into the fact, that
the charge carrier-exchange boson interaction spectral density $\itof{}$ is
reduced at low frequencies due to the onset of superconductivity.
We have already seen in Fig.~\ref{fig:7} the growth of the
$41\,$meV resonance in $\itof{}$ as the temperature is lowered.
At the same time the in-plane infrared optical data shows a gapping
or at least
a strong reduction of spectral weight at small $\omega$. This
implies that for temperatures smaller than the characteristic
energy associated with this reduction, the inelastic scattering
rates will become exponentially small and therefore the
inelastic scattering time will become very large. This feature
by itself will increase the microwave conductivity. At the
same time the normal fluid density is of course decreasing
towards zero. This feature reduces the absorption which is due only
to the normal excitation. The two effects combine to give a
maximum in the real part of the microwave conductivity at
some intermediate temperature.

Another possible way to describe this collapse of the inelastic
scattering time is the introduction of a temperature
dependent inelastic scattering time which can be modeled
from spin fluctuation theory \cite{Hensen}.

\begin{figure}
\vspace*{-4mm}
\includegraphics[width=8cm]{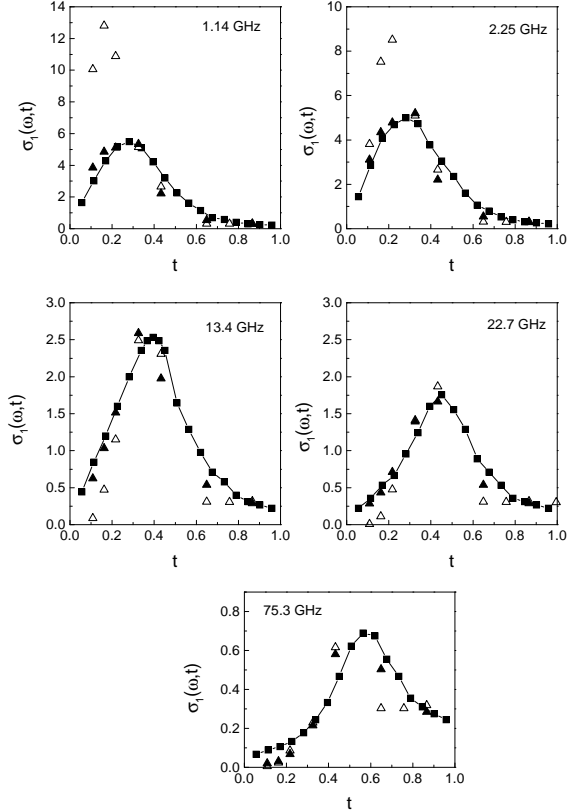}
\vspace*{-7mm}
\caption{
Microwave conductivity $\sigma_1(\omega,t)$ in
$10^7\Omega^{-1}$m$^{-1}$ vs the reduced temperature
$t=T/T_c$ for the five frequencies measured in experimental
work of Hosseini {\it et al.} \protect\cite{hosseini} namely $\Omega =%
1.14$, 2.25, 13.4, 22.7, and $75.3\,$GHz (bottom frame).
Solid squares are experiment, open triangles clean limit and
solid triangles inelastic scattering plus impurities characterized
by a potential with $\Gamma^+ = 0.003\,$meV and $c=0.2$.}
\label{fig:15}
\end{figure}

\begin{figure}
\vspace*{-30mm}
\includegraphics[width=8cm]{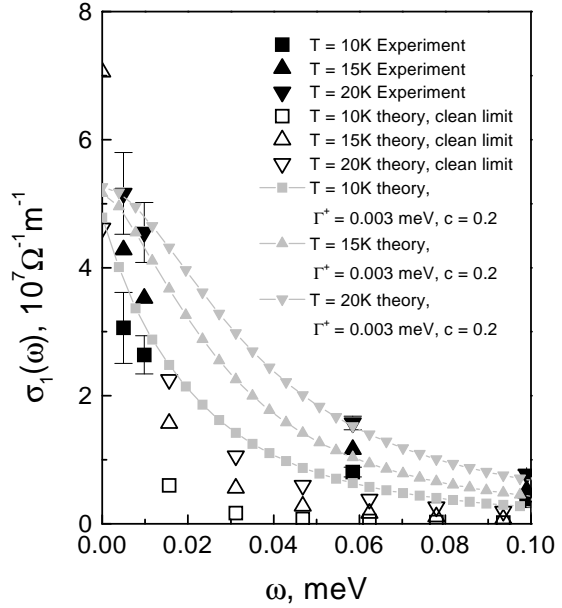}
\vspace*{-8mm}
\caption{The microwave conductivity $\sigma_1(\omega,T)$ as a
function of $\omega$ for three different temperatures. The data
is the same as shown in Fig.~\protect\ref{fig:15}.
The open symbols are theory
for the pure limit, the solid gray symbols theory with some
impurity scattering additionally included, and the solid black
symbols are experiments. The squares are for $T=10\,$K, the
up-triangles for $T=15\,$K, and the down-triangles for
$T=20\,$K.
}
\label{fig:16}
\end{figure}

Recently Hosseini {\it et al.} \cite{hosseini} provided
new microwave data at five different frequencies between 1 and
75 GHz on ultra pure samples of YBa$_2$Cu$_3$O$_{6.99}$ grown
in BaZrO$_3$ crucibles. In Fig.~\ref{fig:15} we show results
obtained from our Eliashberg solutions and compare with
experiment \cite{microw}.
The solid squares are the data of Hosseini {\it et
al.}, the open triangles are our numerical results in the clean limit,
and the solid triangles include a small amount of impurities
characterized by $\Gamma^+=0.003\,$meV and $c=0.2$. The figure
has five frames one for each of the five microwave frequencies
considered, namely 1.14, 2.25, 13.4, 22.7, and $75.3\,$GHz. We
see that even for these ultrapure crystals, results obtained
without including impurities do not agree well with the data
at the lowest microwave frequencies considered and at the
lowest temperatures. For example, in the case of the 
$\omega= 1.14$ and
$2.25\,$GHz runs the predicted peak is much to high. The agreement,
however, improves as the frequency of the microwave probe is
increased. More importantly, when a small amount of impurity
scattering with $c=0.2$ is included, good agreement is obtained
in all cases. The same data plotted in a different way shows
better the dramatic improvement in the agreement with the data when
impurities are included. This is demonstrated in Fig.~\ref{fig:16}
where we show the data for the microwave conductivity
$\sigma_1(\omega)$ vs $\omega$
at three different temperatures. The data are represented
by solid squares, up-triangles and down-triangles
for $T=10\,$K, $15\,$K and $20\,$K respectively.
The open symbols give the results of our Eliashberg
calculations in the pure case and the solid gray symbols include
impurities. The gray lines through the points are a guide for the
eye. The agreement with the data in this last case
is within experimental error and is acceptable. It is clear,
that a small amount of elastic scattering needs to be
included in the calculations to achieve good agreement.

We now turn to the $c$-axis. No new parameters relevant to the
in-plane dynamics need to be introduced
in order to understand the $c$-axis
data. It is necessary, however, to have some model for the
$c$-axis charge transfer. Coherent or incoherent hopping will lead
to quite different conclusions as will the assumption, in the
coherent hopping case, of a constant or a momentum  dependent
hopping probability. For the constant case the conductivity will
mirror its in-plane value but its magnitude will of course be
greatly reduced. For a
$\theta$-dependent matrix element, on the other hand, 
the nodal quasiparticles are eliminated from participation
in the $c$-axis response and we can expect a behavior different
from the in-plane results.

\begin{figure}
\vspace*{-30mm}
\includegraphics[width=8cm]{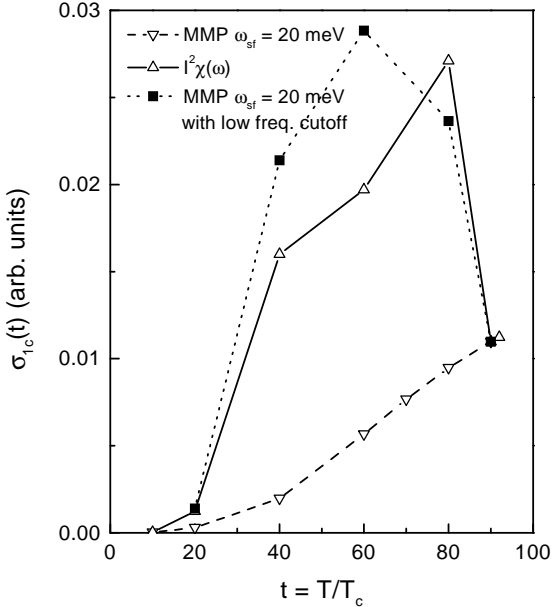}
\vspace*{-10mm}
\caption{
The $c$-axis microwave conductivity $\sigma_{1c}(T)$ at $\omega =
22\,$GHz as a function of temperature $T$. The open up-triangles
were obtained from the empirically determined $I^2\chi(\omega)$
shown in Fig.~\protect{\ref{fig:7}}. The solid squares are for an
MMP form Eq.~(\protect{\ref{eq:2}}) with low frequency cutoff
applied. This cutoff is the same as seen in Fig.~\protect{\ref{fig:7}}.
The open down-triangles employ the same MMP model with $\omega_{SF} =
20\,$meV and without the low frequency cutoff. In this case there
is no peak in $\sigma_{1c}(T)$. All curves are for coherent
tunneling with $t_\perp({\bf k}) = t_{\perp}\cos^2(2\theta)$.
}
\label{fig:17}
\end{figure}

\begin{figure}
\vspace*{-20mm}
\includegraphics[width=8cm]{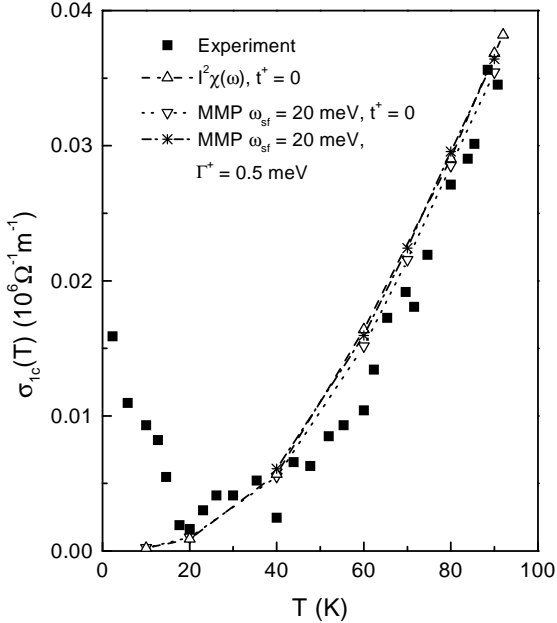}
\vspace*{-18mm}
\caption{
The $c$-axis microwave conductivity $\sigma_{1c}(T)$ at $\omega =
22\,$GHz as a function of temperature $T$. The solid squares are the
experimental results by Hosseini {\it et al.} \protect{\cite{hosseini2}}
shown for comparison. The others are theory based on various models
for the charge carrier-exchange boson interaction spectral density.
All calculations
are for the incoherent case based on Eqs.~(\protect{\ref{eq:4}})
with $V_1 = V_0$ in (\protect{\ref{eq:4c}}) and $V_0$ adjusted to
match the experimental value at $T=90\,$K. Open up-triangles are
based on our empirically determined charge carrier-exchange boson
interaction spectral
density $I^2\chi(\omega)$ of Fig.~\protect{\ref{fig:7}} without
additional impurity scattering (pure limit). The open
down-triangles are also in the pure limit but the MMP model
(\protect{\ref{eq:63}}) is used without cutoff. The stars are the
same as the open down-triangles but now impurity scattering is
included in the unitary limit with $\Gamma^+=0.5\,$meV.
}
\label{fig:18}
\end{figure}

In Fig.~\ref{fig:17} we show numerical results for the temperature
dependence of the real part of the $c$-axis microwave conductivity
$\sigma_{1c}(T)$ at $\omega=22\,$GHz as a function of temperature
$T$ in arbitrary units. The results are for coherent hopping with
a $\theta$-dependent matrix element. For the in-plane case the
same form applies except that the vertex $t_\perp({\bf k})$ would be
replaced by a Fermi velocity $v_{\bf k}$. In as much as both these
vertices are taken to be independent of {\bf k} they can be pulled out
of the integral over {\bf k} in Eqs.~(\ref{eq:2}) and
in-plane and out-of-plane conductivities differ only by a
numerical constant which sets the over all scale in each case.
Since no peak is observed in the temperature dependence of the
$c$-axis conductivity \cite{hosseini2} this case does not agree
with experiment and
will not be treated further here. Only results for
$t_\perp({\bf k}) = t_{\perp}\cos^2(2\theta)$
are considered in Fig.~\ref{fig:17}. The open up-triangles are the
results obtained from the charge carrier-exchange boson
interaction spectral density $\itof{}$
obtained empirically from the in-plane infrared conductivity
(Fig.~\ref{fig:7}). This
is the only material parameter which characterizes YBCO in the
Eliashberg equations (\ref{eq:57}). Solutions of these equations
determine the in-plane Green's function (\ref{eq:4c}) and hence
the $c$-axis conductivity Eqs. (\ref{eq:4}). Arbitrary units are
used, so that the absolute value of $t_{\perp}$ is not required.
We see a broad peak in $\sigma_{1c}(T)$ vs $T$ which is centered
around $T=60\,$K rather than around $T=40\,$K for the in-plane
case of Fig.~\ref{fig:15}. The $c$-axis peak is also smaller. These
differences are entirely due to the extra factor of $\cos^4(2\theta)$
in the $c$-axis conductivity which eliminates the nodal direction.
This has a profound effect on the resulting temperature dependence
of $\sigma_{1c}(T)$ but, as we can see, does not entirely eliminate
the peak in $\sigma_{1c}(T)$. There is considerable disagreement with
experiment which is shown as the solid squares in Fig.~\ref{fig:18}.
Our theoretical results are robust in the sense that the peak is
due to the greatly reduced spectral weight in $I^2\chi(\omega)$
of Fig.~\ref{fig:7} at small $\omega$ when superconductivity sets in
and this is fixed from consideration of the in-plane conductivity.
The effect of this spectral weight reduction is further illustrated
in Fig.~\ref{fig:17} by the solid squares which employ instead of
our empirical value for $I^2\chi(\omega)$ the simpler MMP form
of Eq.~(\ref{eq:63}) with the same low frequency cutoff as indicated
in Fig.~\ref{fig:7} being applied. The cutoff is of course temperature
dependent and goes to zero at $T_c$.
The peak in $\sigma_{1c}(T)$ vs $T$ remains and is
close to the results obtained when an optical resonance is
included in addition to a low frequency cutoff. For comparison,
the down-triangles were obtained when no low frequency cutoff was
applied. We now see that the peak in $\sigma_{1c}(T)$ vs $T$ is
completely eliminated. This demonstrates that the peak is due
to the collapse of the inelastic scattering rate embodied in
the low frequency gapping of the charge carrier-boson spectral
density. In summary, even when a momentum dependent coherent
hopping matrix  element of the form $t_\perp({\bf k}) =%
t_{\perp}\cos^2(2\theta)$
 is considered, gaping of $I^2\chi(\omega)$
at small $\omega$ leads directly to a peak in the $c$-axis
microwave conductivity. However, the spectral density
$I^2\chi(\omega)$ which enters the Eliashberg equations
 (\ref{eq:ImagEli})
on the imaginary axis and (\ref{eq:57}) on the real axis could
depend on  position on the Fermi surface.
This complication was not considered here
but it is important to point out that coherent $c$-axis
tunneling could lead to reasonable agreement with the measured
temperature variation of the microwave conductivity, if the
spectral density $I^2\chi(\omega)$ is different along the
antinodal direction and, in particular, has no gapping at low
frequencies. This difference might not have been picked up in
our analysis of the in-plane conductivity which is characteristic of
an average over all points on the Fermi surface and not just
of the antinodal direction.

Fig.~\ref{fig:18} shows results for the temperature variation of the
$c$-axis microwave conductivity $\sigma_{1c}(T)$ at
$\omega=22\,$GHz in the incoherent coupling case, Eqs.~(\ref{eq:4}).
As we have indicated in Sec.~2 this formula involves a double
integral over momentum which separately weights the two Green's
functions. For simplicity, we show results only for the case
$V_1 = V_0$ in the impurity model potential of Eq.~(\ref{eq:4c}).
Other values have been considered but this does not change
qualitatively any of the conclusions we will make. The open
up-triangles give results when the empirical $I^2\chi(\omega)$
of Fig.~\ref{fig:7} is used. We see that in this case the theory
predicts no peak in $\sigma_{1c}(T)$ vs $T$ in good agreement
with the experimental results of Hosseini {\it et al.}
\cite{hosseini2} (solid squares). The low frequency cutoff
built into our $I^2\chi(\omega)$ (open up-triangles)
has little effect on the
resulting $\sigma_{1c}(T)$. This is verified directly when we
compare with the open down-triangles which were obtained with
the MMP form (\ref{eq:63}) without cutoff. These results differ
very little from the previous ones and show that the application
of a low frequency cutoff does not play a critical role for the
incoherent case. This is in sharp contrast to the coherent case
in which the low frequency cutoff leads directly to a peak in
$\sigma_{1c}(T)$. The curves are also robust to
the introduction of some elastic impurity scattering as is
demonstrated with
the final set of results in Fig.~\ref{fig:18}, denoted by
stars, which is based on an MMP model with elastic impurity scattering
included in the unitary limit with $\Gamma^+=0.5\,$meV in Eq.~%
(\ref{eq:ImagEliD}). We see that the inclusion of impurities does not
appreciably change our results. The calculation clearly shows
that the observed data can be understood naturally in an
incoherent $c$-axis transport model and that the results are
robust to changes in cutoff at low $\omega$ and to the
addition of impurities.

This contrasts with the case of the $c$-axis infrared data which
we described previously and found to support coherent rather
than incoherent $c$-axis charge transfer.

\subsubsection{Other Superconducting State Properties}

\begin{figure}
\vspace*{-2mm}
\includegraphics[width=8cm]{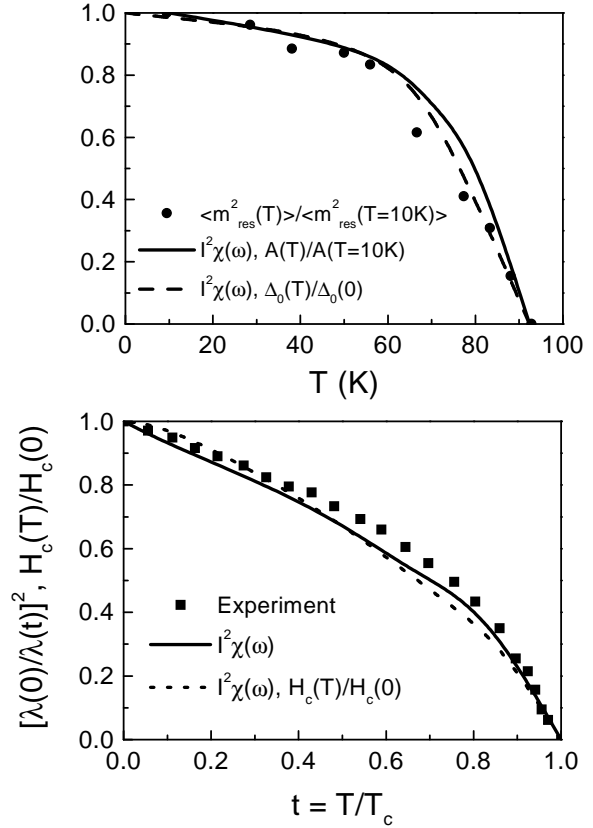}
\vspace*{-4mm}
\caption{
Top frame: spectral weight under the optical resonance as a
function of temperature (solid line) obtained from the
optical data of Fig.~\protect\ref{fig:6} (top frame, five temperatures
only). The solid circles are the data of Dai {\it et al.}
\protect\cite{Dai} for the normalized area under the spin resonance
 obtained by neutron scattering.  The
dashed curve gives our calculated $\Delta_0(T)/\Delta_0(0)$.
Bottom frame: the normalized London penetration depth squared
$(\lambda(0)/\lambda(T))^2$ vs reduced temperature $t = T/T_c$
(solid line) compared with
the experimental results of Bonn {\it et al.} \protect\cite{Bonn}.
The dotted curve gives
thermodynamic critical field $H_c(T)/H_c(0)$ vs $t$.}
\label{fig:19}
\end{figure}

The temperature dependence of the area under the spin resonance
seen at $(\pi,\pi)$ by spin polarized neutron scattering \cite{Dai}
has been measured and its temperature
dependence denoted by $\langle m^2_{res}(T)\rangle/\langle%
m^2_{res}(T=10\,{\rm K})\rangle$ is reproduced in the top frame
of Fig.~\ref{fig:19} as the solid circles. Also shown on the
same plot are our results for the area under the optical resonance
in our spectral density $\itof{}$ at various temperatures
(see Fig.~\ref{fig:7}).
We denote this by $A(T)$ and plot as the solid line the ratio
$A(T)/A(T=10\,{\rm K})$ which follows the same temperature
variation as the neutron result. This temperature variation
is also close to that of the gap edge shown as the dashed
curve. This last curve was found as a byproduct of our Eliashberg
calculations based on the numerical solutions of Eqs.~%
(\ref{eq:57}). The gap amplitude at temperature $T$ is given by
$\Re{\rm e}\{\Delta(\omega=\Delta_0;T)\} = \Delta_0(T)$.
The same solutions
give the temperature dependence of the penetration depth and
of the thermodynamic critical field which we present in the
bottom frame of Fig.~\ref{fig:19}. The solid curve gives results
for $[\lambda(0)/\lambda(T)]^2$ vs $t = T/T_c$ (the reduced
temperature) which are close to the experimental results of
Bonn {\it et al.} \cite{Bonn} given as solid squares.

\begin{figure}
\vspace*{-3mm}
\includegraphics[width=8cm]{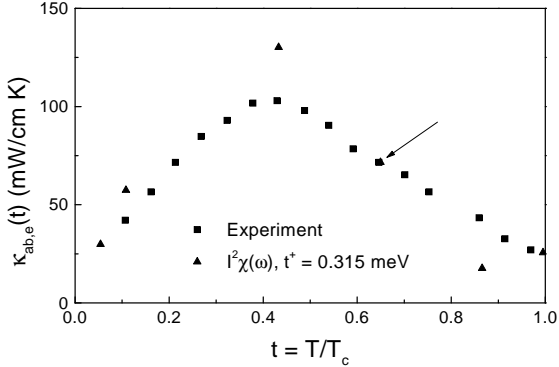}
\vspace*{-5mm}
\caption{
The electronic part of the thermal conductivity as a function
of the reduced temperature $t = T/T_c$. The solid squares
are the experimental results of Matsukawa {\it et al.}
\protect\cite{matsukawa} and the solid triangles our theoretical result
with some impurity scattering included.
\label{fig:20}}
\end{figure}

Further results are shown
in Fig.~\ref{fig:20} for the electronic part of the thermal
conductivity as a function of temperature. It shows, as does the in-plane
microwave conductivity, a large peak around $40\,$K. The
solid triangles are the results of our calculations while
the solid squares are the experimental results of Matsukawa
{\it et al.} \cite{matsukawa}. The agreement is remarkably
good. We stress that no adjustable parameters enter our
calculations except for a choice for the impurity parameter
$t^+$ which is also restricted in this particular case
because $t^+$ has already been determined by a previous fit to the
microwave data \cite{schach2}.
The kernel in the Eliashberg equations is completely determined
from optical data. 

Other important successes of our Eliashberg calculations are
summarized in Table~\ref{t1}. We begin with a discussion of the
plasma frequency $\Omega_p$. Referring to Fig.~\ref{fig:9} we
point out the arrow which shows the frequency at which we made
our calculated conductivity agree exactly with experiment. This
sets the plasma frequency which is also the total spectral weight
under the real part of the conductivity. The optical spectrum
sum rule is
\be
 \ili_0^\infty\!d\omega\,\sigma_1(\omega) = {\Omega^2_p\over 8}.
\ee
A value of $\Omega_p=2.36\,$eV is found which agrees well with
the experimental value $2.648\,$eV (see Tab.~\ref{t1}).
\begin{table}[t]
\caption{Some superconducting properties of the twinned YBCO sample:
$\Delta F(0)$ is the condensation energy at $T=0$ in meV/Cu-atom,
$n_s/n$ is the superfluid to total carrier density ratio, $\Omega_p$ is
the plasma frequency in eV.}
\label{t1}
\begin{ruledtabular}
\begin{tabular}{cccc}
 & Theory & Experiment & Ref. \\
\hline
$\Delta F(0)$ & 0.287 & 0.25 & \protect\cite{Norm,Loram} \\
$n_s/n$ & 0.33 & 0.25 & \protect\cite{Tanner} \\
$\Omega_p$ & 2.36 & 2.648 & \protect\cite{Tanner2} \\
$2\Delta_0/k_BT_c$ & 5.1 & 5.0 & \protect\cite{Dynes} \\
\end{tabular}
\end{ruledtabular}
\end{table}
A further comparison of our model with the infrared data is provided
by the analysis of the fraction of the total normal-state spectral
weight which condenses into the superfluid: $n_s/n$. Indeed, strong
electron-boson coupling reduces the spectral weight of the
quasiparticle component of the electronic spectral function
$A({\bf k},\omega)$ compared to its non-interacting value by
a factor of $Z$ leading at the same time to the appearance of an
incoherent component. It is the latter component which is
responsible for the Holstein band in the optical conductivity
whereas the coherent  quasiparticle part gives rise to the
Drude term at $T>T_c$ and to the superfluid density at $T=0$ in
the spectra of $\sigma_1(\omega)$ \cite{footn1}. The values
of $n_s/n$ and hence $(Z-1)$ yield an estimate of the strength of
renormalization effects in the interacting system.
Tanner {\it et al.} \cite{Tanner} obtained $n_s/n\simeq 0.25$
in crystals of YBCO. This compares well with the value
$\simeq 0.33$
which corresponds to $Z\simeq 3$ (at low temperatures) generated
in our analysis.

We have also calculated the condensation energy \cite{Carb1} as a function
of temperature. Its value at $T=0$
follows from the normal-state electronic density of states which we take
from band structure theory equal to $2.0\,$states/eV/Cu-atom
(double spin) around the middle of the calculated range of values
\cite{Junod}.
This gives a condensation energy $\Delta F(0) = 0.287\,{\rm meV/Cu-atom}$
which agrees well with the value quoted by Norman {\it et al.} \cite{Norm}
from the work by Loram {\it et al.} \cite{Loram}. (See Tab.~\ref{t1}.)
This is equivalent to a thermodynamic critical field
$\mu_0 H_c(0) = 1.41\,$T with $H_c(T)$ defined through
$\Delta F(T) = H^2_c(T)/8\pi$. The normalized value $H_c(T)/H_c(0)$ is shown
as the dotted line in the bottom frame of Fig.~\ref{fig:19}
and is seen to follow reasonably, but not exactly, the $T$
dependence of the normalized penetration depth.
One further quantity is the ratio of the gap amplitude
to the critical temperature which in BCS theory is
$2\Delta_0/k_BT_c = 4.2$ for $d$-wave. In Eliashberg theory the gap depends
on frequency. In this case an unambiguous definition
of what is meant by $\Delta_0$ is to use the position in energy
of the peak in the quasiparticle density of states which
is how the gap $\Delta_0$
is usually defined experimentally for a $d$-wave superconductor. We get
a theoretical value of $ 2\Delta_0/k_BT_c \simeq 5.1$ in good agreement
with experiment, as shown in Tab.~\ref{t1}.

\subsection{The Compound Bi$_2$Sr$_2$CaCu$_2$O$_{8+\delta}$}

Optical data published by Puchkov {\it et al.} \cite{puchkov}
\begin{figure}
\vspace*{-24mm}
\includegraphics[width=8cm]{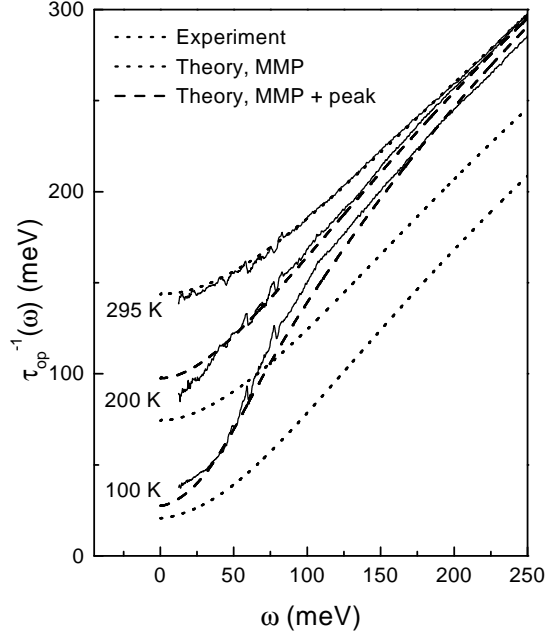}
\vspace*{-1cm}
\caption{The temperature dependent $ab$-plane optical scattering
rate for an optimally doped Bi2212 single crystal with $E||a$ in
the normal state. The solid lines represent experimental data by
Tu {\it et al.} \protect{\cite{Tu}}. The dotted lines give the theoretical
result calculated using Eq.~(\protect{\ref{eq:42}}) and a $\itof{}$
which is just an MMP spectrum
with $\omega_{SF} = 82\,$meV. At $T=295\,$K theory reproduces
experiment almost ideally; this agreement deteriorates at lower
temperatures. Finally, the dashed lines present theoretical results
found for the same MMP spectrum as before but now with coupling
to the $43\,$meV optical resonance added.}
\label{fig:21}
\end{figure}
for optimally doped samples of the compound
Bi$_2$Sr$_2$CaCu$_2$O$_{8+\delta}$ (Bi2212) have first been
analyzed by Schachinger and Carbotte \cite{schach6}. They reported
that the normal-state optical scattering rate ($T=300\,$K) can
be fitted perfectly by an MMP spectrum with $\omega_{SF} = 100\,$meV
and with an high energy cutoff at $400\,$meV.
The inversion of the superconducting state optical scattering rate
revealed the coupling of the charge carriers to a resonance found
at an energy of $43\,$meV. This corresponds to the magnetic resonant
mode found by Fong {\it et al.} \cite{Fong} using inelastic neutron
scattering. This mode appears below $T_c$ and its intensity increases
with decreasing temperature.

\begin{figure}
\vspace*{-37mm}
\includegraphics[width=8cm]{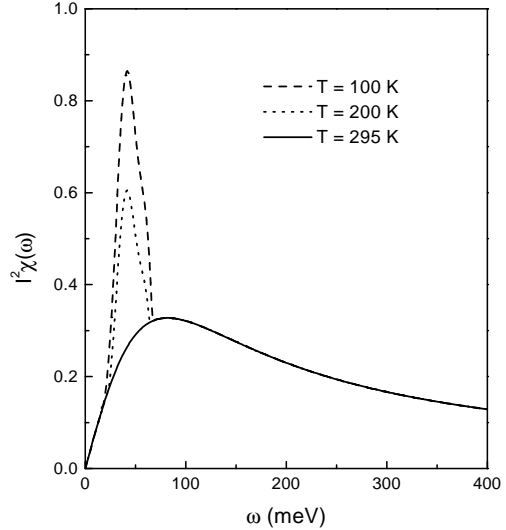}
\vspace*{-10mm}
\caption{The charge carrier-spin excitation spectral density
$\itof{}$ determined from normal-state optical scattering data
shown in Fig.~\protect{\ref{fig:21}} for optimally doped Bi2212
single crystals. The solid curve is for $T=295\,$K, the dotted
curve for $200\,$K, and the dashed one for $100\,$K. Note the
growth in strength of the $43\,$meV optical resonance as the
temperature is lowered.}
\label{fig:22}
\end{figure}

Tu {\it et al.} \cite{Tu} recently studied the $ab$-plane
charge dynamics in
optimally doped Bi2212 single crystals ($T_c = 91\,$K)
using an experimental technique with much
improved signal to noise ratio. They developed an experimentally
unambiguous method which examines the maxima and minima of
$W(\omega)$, Eq.~(\ref{eq:47}). The authors argued that a comparison
of their spectral data with data found for YBCO suggests that a
pseudogap exists in Bi2212 above $T_c$, at least at $T=100\,$K.
Fig.~\ref{fig:21} presents their data (solid lines) together with
a theoretical analysis. For $\itof{}$ an MMP spectrum is used and
at $T=295\,$K our best fit is found for $\omega_{SF} = 82\,$meV
together with an high energy cutoff $400\,$meV
using theoretical results calculated from Eq.~(\ref{eq:42}) (dotted
line). If we calculate the optical scattering rate for the
temperatures $T=200\,$ and $100\,$K using the same $\itof{}$
spectrum it becomes obvious that the agreement with experiment
deteriorates with decreasing temperature (dotted lines). A
comparison of the anomaly in the optical data at $T=100\,$K
around $50\,$meV with the optical data of YBCO (Fig.~\ref{fig:8})
reveals quite similar behavior which suggests that in Bi2212
a coupling to an optical resonance can actually be seen in the
normal state. Indeed, the quality of the data is good enough
to allow us to derive $W(\omega)$ by inversion. The result shows
a pronounced peak at $43\,$meV which can be used to modify the
$\itof{}$ spectrum. This is shown in Fig.~\ref{fig:22}.
Using this modified spectrum $\itof{}$ to calculate the optical
scattering rate for $T=100\,$K results in excellent agreement
between experiment and theory (dashed line, Fig.~\ref{fig:21}
labeled $100\,$K).

The $T=200\,$K data show a similar, but less pronounced, anomaly.
It is not possible to derive a $W(\omega)$ directly from the data so
we simply use the $\itof{}$ found for $100\,$K and reduce the
size of the $43\,$meV peak until best agreement between experiment
and theory is reached (dashed line in Fig.~\ref{fig:21} labeled
$200\,$K). This results in an $\itof{}$ presented
in Fig.~\ref{fig:22} which still contains a pronounced contribution
from the coupling of the charge carriers to the optical resonance
(dotted line).
From this we can conclude that the optical resonance, and probably
connected with it, the magnetic resonant mode exists at least up to
$200\,$K.
Nevertheless, our result for $295\,$K indicates that at this
temperature the optical resonance no longer exists.

\begin{figure}
\vspace*{-3mm}
\includegraphics[width=8cm]{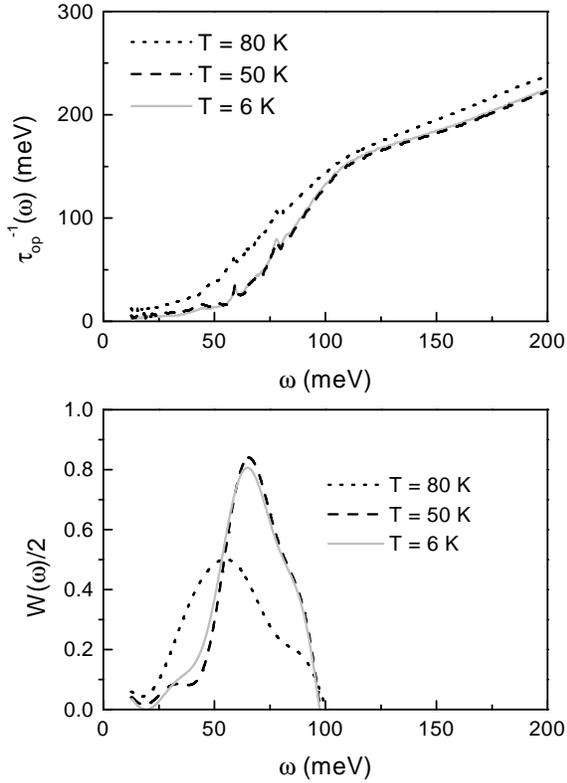}
\vspace*{-10mm}
\caption{
Top frame: optical scattering rate $\tau^{-1}_{op}(T,\omega)$ in
meV for optimally doped, untwinned Bi2212 single crystals
\protect{\cite{Tu}}. Bottom
frame: function $W(\omega)/2$ vs $\omega$ in the region of the
optical resonance.} 
\label{fig:23}
\end{figure}

We will now concentrate on the superconducting state and study
the temperature dependence of $\itof{}$ below $T_c$. The top
frame of Fig.~\ref{fig:23} presents the infrared scattering
rate as measured by Tu {\it et al.} \cite{Tu} for three
temperatures, namely $6\,$K (gray solid line), $50\,$K (dashed
line), and $80\,$K (dotted line). In comparison with similar
results for YBCO (top frame of Fig.~\ref{fig:6}) we recognize
that even at $80\,$K Bi2212 shows a very strong suppression
of $\tau_{op}^{-1}(\omega)$ at energies below $50\,$meV which
is an indication of stronger coupling of the charge carriers
to the optical resonance. The bottom frame of this figure
shows the function $W(\omega)/2$ derived from experiment with
the high energy negative parts suppressed because we want to
concentrate on the optical resonance. It increases as $T$ is
lowered and shows only little further variation below $50\,$K.
In $W(\omega)/2$ the resonance peak is positioned at the
resonance energy $E_r$ plus the gap value $\Delta_0(T)$ and
with the temperature dependence of the gap accounted for, we
can conclude that the position of the resonance is temperature
independent and stays at $E_r=43\,$meV, the energy at which
the magnetic resonant mode is found by inelastic neutron
scattering \cite{Fong}. The coupling of the charge carriers to a boson
at $43\,$meV has also been observed in photoemission 
\cite{Johnson} and tunneling \cite{Zasad} work on Bi2212
and the $43\,$meV magnetic resonant mode seems to be the
obvious candidate for the origin of this boson.

\begin{figure}
\vspace*{-5mm}
\includegraphics[width=8cm]{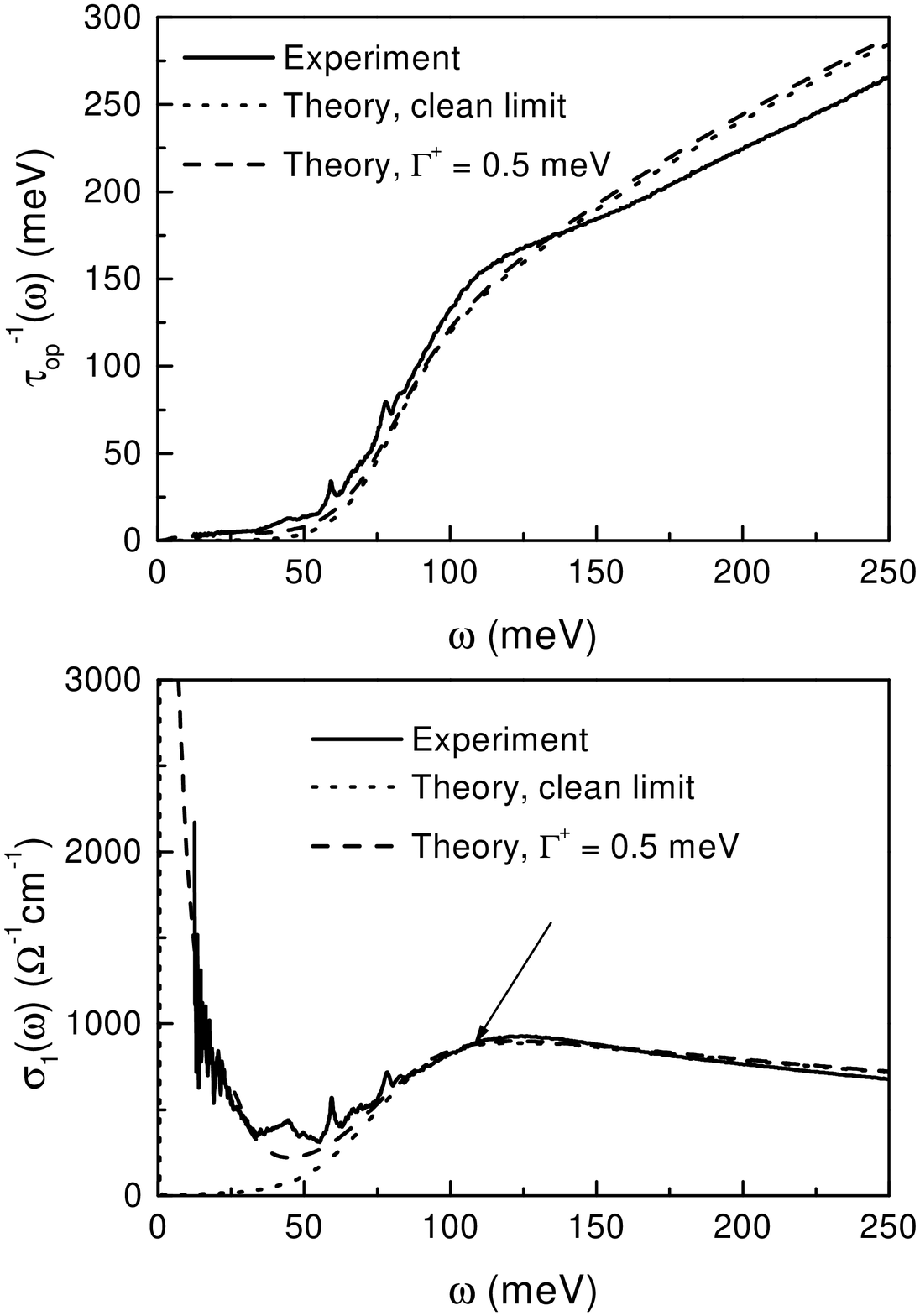}
\vspace*{-4mm}
\caption{
The top frame gives
the optical scattering rate $\tau^{-1}_{op}(\omega)$ vs
$\omega$ for optimally doped Bi2212 single crystals at a temperature
of $6\,$K. The experimental results give the solid curve and
our clean limit theoretical fit to it is the dotted curve. The
dashed curve presents theoretical results for a system with
impurity scattering in the unitary limit described by the
parameter $\Gamma^+ = 0.5\,$meV and $c=0$. The lower frame gives
a comparison of the real part of the in-plane optical conductivity
$\sigma_1(\omega)$ vs $\omega$ for the two models already presented
in the top frame. The solid curve is the experimental data, the
dotted curve the clean limit theoretical result, and the dashed
curve the theoretical result for the system with impurities.}
\label{fig:25}
\end{figure}

Fig.~\ref{fig:25} demonstrates the agreement which can be
achieved between theory and experiment.
The top frame shows the infrared scattering
rate $\tau_{op}^{-1}(\omega)$ vs $\omega$ for $T=6\,$K. The
solid line gives the experimental data while the dashed and
dotted curves represent theoretical results for a clean limit
system and for a system with impurity scattering in the
unitary limit ($\Gamma^+ = 0.5\,$meV) respectively. The differences
in the scattering rate are marginal for these two model systems,
nevertheless, they become important when the real part of the
optical conductivity $\sigma_1(\omega)$ is investigated. The
bottom frame of Fig.~\ref{fig:25} shows the results. The solid
line is experiment, the dotted line is theory for the clean
limit system. It reproduces nicely the maximum in $\sigma_1(\omega)$
around $120\,$meV and the high energy tail. At energies below
$75\,$meV the clean limit results deviates strongly from
experiment towards nearly zero values and show a very pronounced, narrow
peak around $\omega=0$. Results for the system with impurities
treated in the unitary limit ($\Gamma^+ = 0.5\,$meV, dashed line)
display all important features observed in the experimental data.
We see, as in the case of YBCO, that impurities affect only the
low energy region ($\omega < 60\,$meV), the region
$60 \le\omega\le100\,$meV is dominated by the coupling to the
optical resonance modeled in the $\itof{}$ while the energy region
$\omega > 120\,$meV is determined by the normal-state MMP part
of $\itof{}$ as has already been described for the YBCO compound.

The bottom frame of Fig.~\ref{fig:25} contains an arrow which
points out the frequency at which we made our calculated
$\sigma_1(\omega)$ to agree exactly with experiment. (This was
\begin{figure}
\vspace*{-30mm}
\includegraphics[width=8cm]{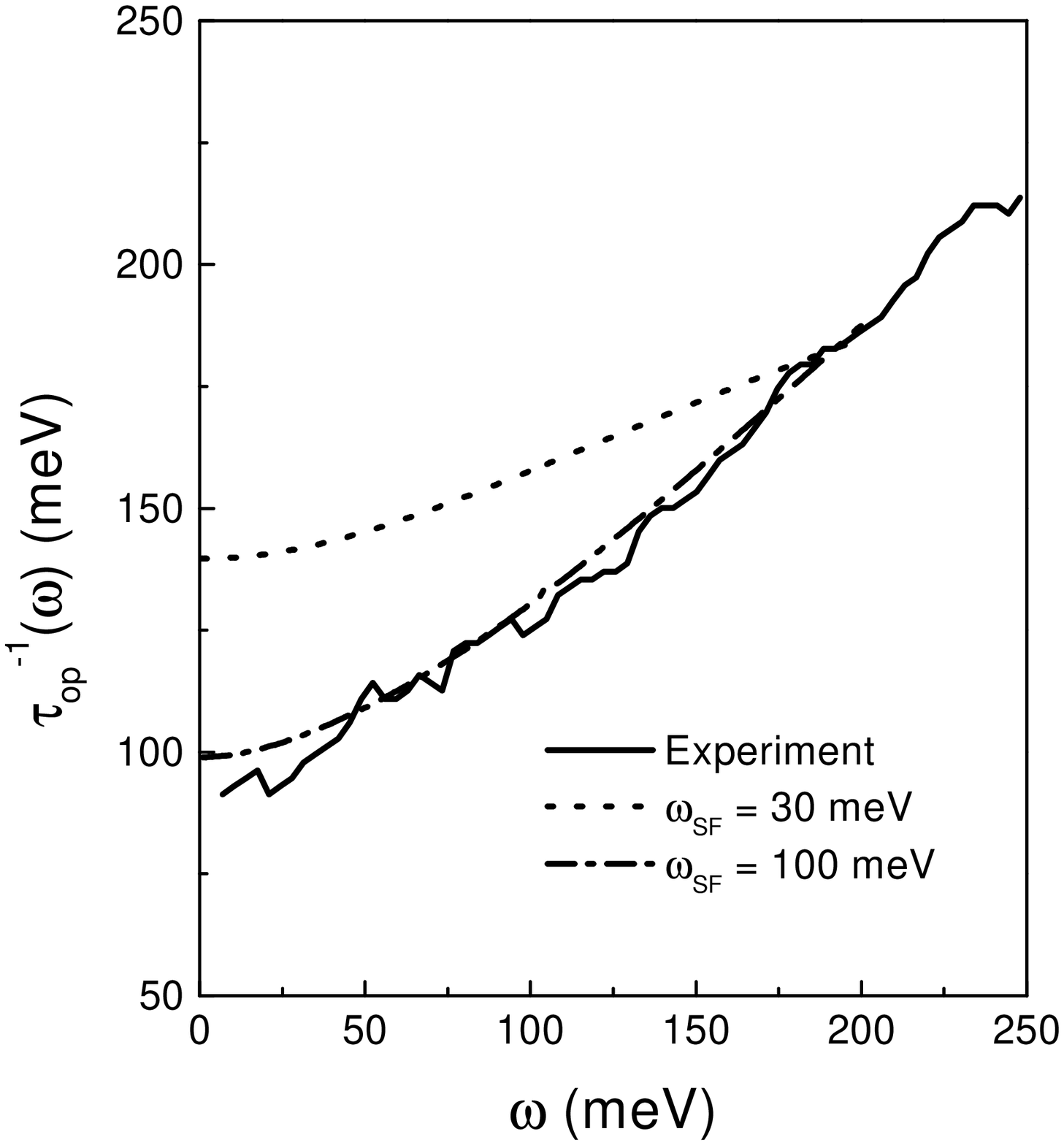}
\vspace*{-13mm}
\caption{
The normal-state optical scattering rate $\tau_{op}^{-1}(\omega)$
vs $\omega$ for Tl2201 with a $T_c = 90\,$K obtained from the
work of Puchkov {\it et al.} \protect\cite{puchkov} (solid curve). The
dash-dotted curve from theory based on Eq.~(\protect\ref{eq:42}) with
an MMP model spectral density using a spin fluctuation frequency
$\omega_{SF} = 100\,$meV gives good agreement while the choice
of $30\,$meV (dotted curve) does not.
\label{fig:26}}
\end{figure}
only done for the clean limit calculation, the same scaling was
used for the system with impurities.) This sets the plasma
frequency $\Omega_p = 2.3\,$eV which is to be compared with
the $\Omega_p = 1.98\,$eV used by Tu {\it et al.} \cite{Tu}.
Finally, we found for the superconducting gap at $T=6\,$K a
value of $25\,$meV. This is certainly smaller than the value
of $34\,$meV reported by R\"ubhausen {\it et al.} \cite{Rub}
from Raman spectroscopy on Bi2212 single crystals with a
$T_c$ of $95\,$K.

\subsection{Application to Other Cuprates}

In contrast to the systems studied so far Tl$_2$Ba$_2$CuO$_{6+\delta}$
(Tl2201) is a monolayer compound while YBCO, Bi2212 and
YBa$_2$Cu$_4$O$_8$ (Y124) are bilayer compounds. Moreover,
Tl2201 is the only system with tetragonal symmetry,
all the other compounds are of orthorhombic symmetry. Its
$T_c\sim 90\,$K and this is similar to the $T_c$ of YBCO and
Bi2212. Y124, on the other hand has a slightly lower $T_c$ of $82\,$K
and shows properties which resemble a moderately underdoped YBCO
compound.

In Fig.~\ref{fig:26} we show our result for the normal-state
$\tau_{op}^{-1}(\omega)$ \cite{puchkov} related to the
conductivity by Eq.~(\ref{eq:46}) for
Tl2201 with $T_c=90\,$K at temperature $T=300\,$K. The solid
curve is the data of Puchkov {\it et al.} \cite{puchkov}. The
dotted curve is our best fit for $\omega_{SF} = 30\,$meV with
$I^2$ adjusted to get the correct absolute value of the
scattering rate at $T=300\,$K and $\omega=200\,$meV. We see that
this value of
$\omega_{SF}$ does not give a satisfactory fit to the data.
The dash-dotted curve, however, fits the data well and
corresponds to $\omega_{SF} = 100\,$meV. This fit provides
us with a model $\itof{}$ valid for the normal state of Tl2201.
This $\itof{}$ is then used to calculate the anisotropy parameter
$g$ from the solution of the linearized imaginary axis Eliashberg
equations (\ref{eq:ImagEli}) for the critical temperature
$T_c = 90\,$K. As a result of this procedure all necessary parameters
are fixed and we can now proceed to study the superconducting
state.

\begin{figure}
\vspace*{-5mm}
\includegraphics[width=8cm]{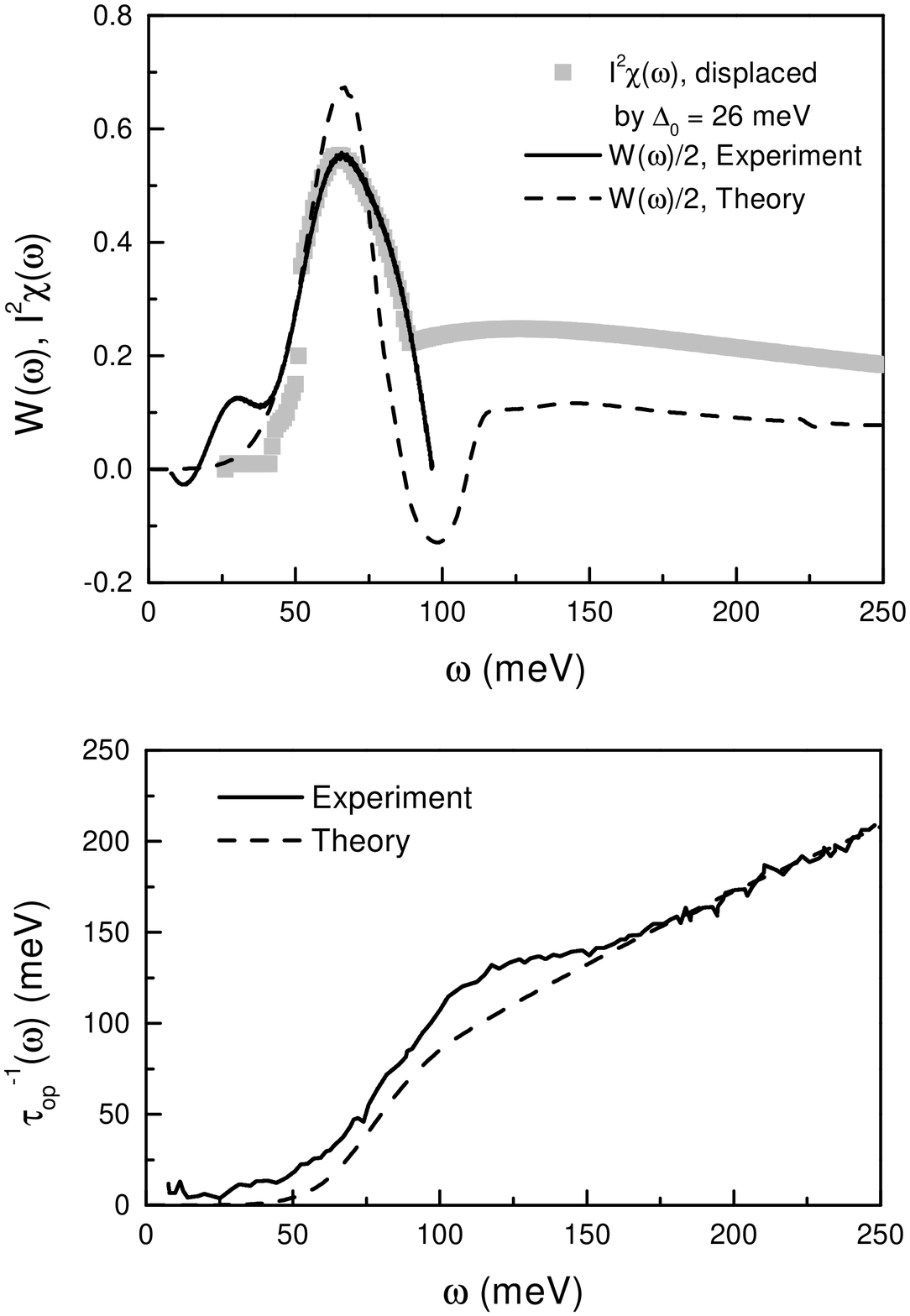}
\vspace*{-8mm}
\caption{The top frame gives our model for the spin-fluctuation
spectral density (displaced by the theoretical gap $\Delta_0 =
26\,$meV for Tl2201 in the superconducting state at $T=10\,$K
(gray solid squares). The dashed line is $W(\omega)$ obtained
from the calculated conductivity and the black solid line
is the coupling to the resonance found directly from experiment.
(The high frequency part has been omitted.)
It was used to in constructing the model $I^2\chi(\omega)$.
The bottom frame shows the optical scattering rate at
$T=10\,$K (solid line) and the theoretical fit to experiment
found from Eliashberg theory.}
\label{fig:27}
\end{figure}

Results are shown in Fig.~\ref{fig:27}. The solid line in the
top frame shows the optical resonance obtained from inversion
of the experimental superconducting optical scattering rate,
presented in the bottom frame of this figure (solid line).
The gray squares are the $\itof{}$
used in the calculations displaced in energy by the gap
$\Delta_0 = 26\,$meV. It is constructed completely from experiment
and we followed the procedure already described in detail for
the YBCO compound. The dashed curve, finally, is the result of
an inversion of theoretical data, shown in the bottom frame
of this figure (dashed line) and we see that it agrees reasonably
well with experiment (solid line). These results allowed Schachinger and
Carbotte \cite{schach5} to predict for Tl2201 a spin resonance
at $43\,$meV. They also predicted that the resonance should
be less pronounced and broader in Tl2201 than in YBCO or Bi2212.
Recently He {\it et al.} \cite{He} succeeded in preparing
a Tl2201 sample big enough for inelastic magnetic neutron
scattering. This sample consists of about 300 coaligned optimally
doped Tl2201 single crystals. This experiment confirmed the
existence of a magnetic resonant mode in Tl2201
below $T_c$ which is located
at about $47\,$meV and which appears to be narrower than the
resonances observed in YBCO or  Bi2212. This is in slight
disagreement with the results of Schachinger and Carbotte 
\cite{schach5} and
this disagreement could probably be explained by the poorer
quality of the samples used by Puchkov {\it et al.} \cite{puchkov}
for the optical measurements many years ago. Other optical data
are not available. Nevertheless, the basic agreement between
the observation of Schachinger and Carbotte \cite{schach5} that
the charge carriers in Tl2201 couple to an optical resonance and
the subsequent observation of a magnetic resonant mode at about
the same energy by He {\it et al.} \cite{He} using inelastic
neutron scattering is quite important. It proves that magnetic
resonant modes are not restricted to bilayer compounds and that
we seem to be confronted with a unified phenomenological picture.

\begin{figure}
\vspace*{-17mm}
\includegraphics[width=8cm]{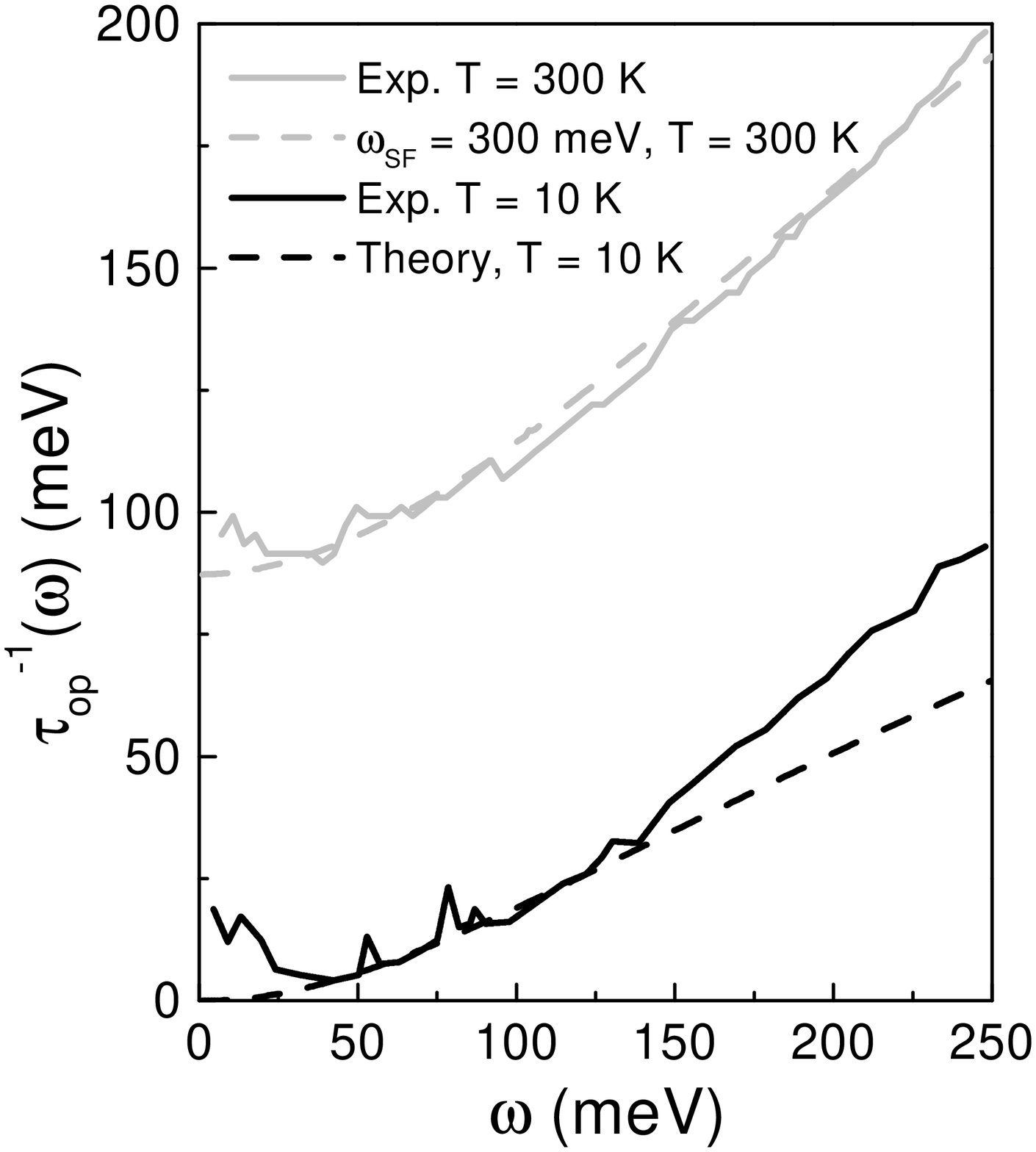}
\vspace*{-20mm}
\caption{
The optical scattering rates in an overdoped sample of Tl2201
with a $T_c=23\,$K. The solid lines represent experimental data
and the dashed lines fits. The gray curves apply in the normal-%
state at $T=300\,$K and the black curves in the superconducting
state at $T=10\,$K. No optical resonance peak is found in this
case.
\label{fig:28}}
\end{figure}

The optical resonance peak is not observed in all systems as is
illustrated in Fig.~\ref{fig:28} for an overdoped sample of Tl2201
with $T_c = 23\,$K. In this case a fit to the $T=300\,$K normal-%
state data (solid gray curve) with an MMP model
gives $\omega_{SF} = 100\,$meV (gray dashed curve). The same
spectrum also produces a good fit (black dashed line) to
the data at $T=10\,$K (black solid line) in the superconducting
state. There is no need to introduce a spin resonance. Indeed
the black solid curve for the measured optical scattering rate
$\tau_{op}^{-1}(\omega)$ is smooth and increases gradually as
$\omega$ increases with no clear sharp rise at any definite
frequency in sharp contrast with Fig.~\ref{fig:27}.
We conclude from this analysis that the resonance observed
in some cuprates with high values of $T_c$ at optimum doping is
not present in all cases and in particular there is no evidence
for such a resonance in overdoped Tl2201 with $T_c = 23\,$K. In
this case a standard MMP spectrum of the form (\ref{eq:63})
gives an adequate representation of the superconducting state optical
scattering rate as a function of $\omega$ with the same
spectral density as was determined by
the data at $T=300\,$K. This is
in contrast to the other cases studied above for which the onset
of superconductivity appears to produce essential modifications of
the underlying spectral density $\itof{}$.

\begin{figure}
\vspace*{-5mm}
\includegraphics[width=8cm]{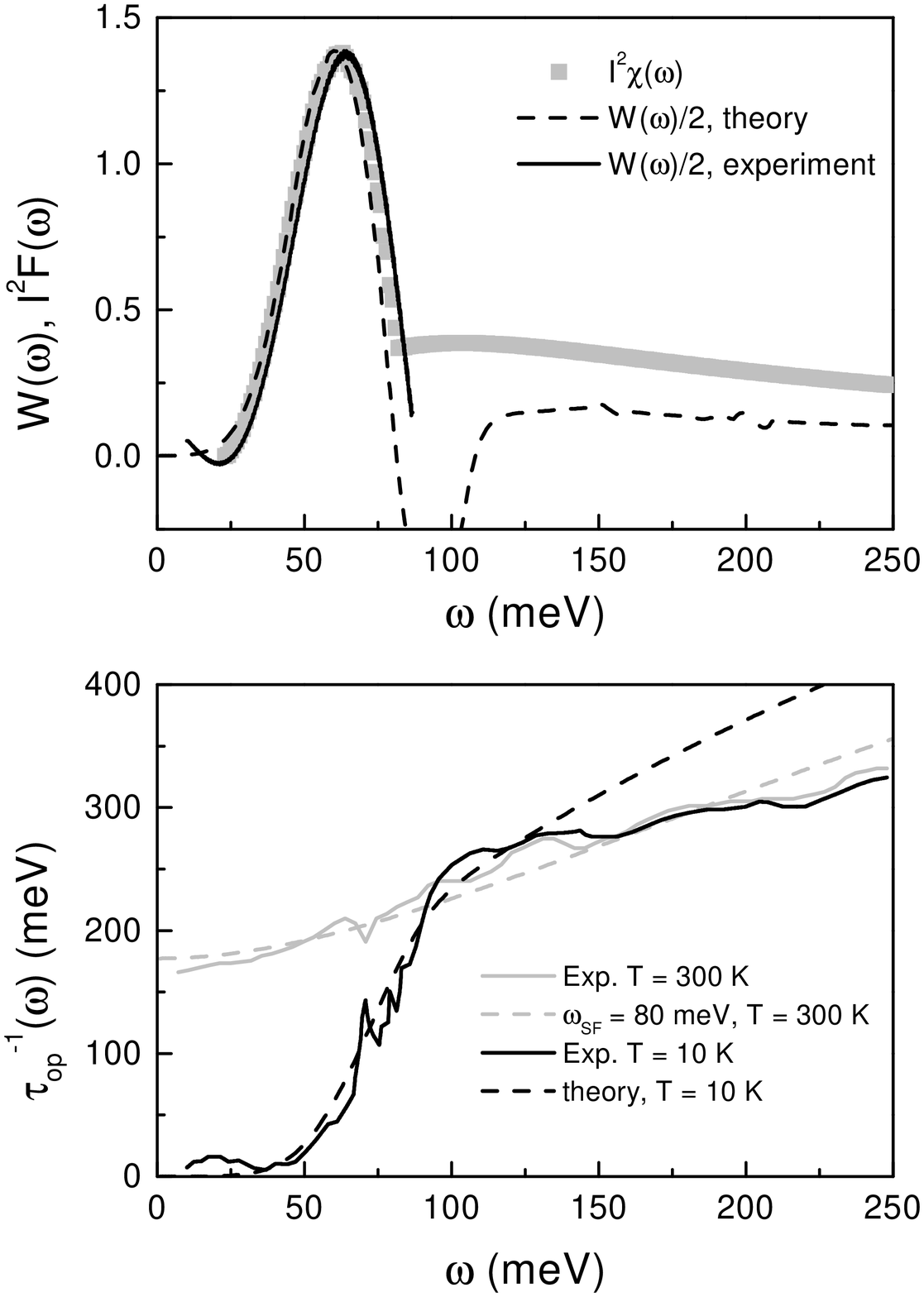}
\vspace*{-5mm}
\caption{
The same as Fig.~\protect{\ref{fig:27}} but for the material
Y124. The spin-fluctuation spectral density $\itof{}$ was
displaced by the theoretical gap $\Delta_0=24\,$meV in the
top frame. In addition, the grayed lines in the bottom frame
of this figure show the comparison between experimental and
theoretical normal-state data at $T=300\,$K. Due to this
comparison $\omega_{SF} = 80\,$meV for the MMP spectrum of
Eq.~(\protect{\ref{eq:63}}).}
\label{fig:29}
\end{figure}

We extend our analysis to the material Y124 $(T_c=82\,$K) where
we predict from Fig.~\ref{fig:29} (top frame, solid curve) a
resonance at $38\,$meV. (This is below the energy of $41\,$meV
for the resonance in YBCO which is not surprising as it is
a well established property of bilayer high-$T_c$ cuprates
that the energy of the magnetic resonant mode tracks $T_c$
in underdoped systems \cite{Dai} and the optical resonance
seems to be closely related to this magnetic resonant mode.)
The top frame of this figure demonstrates
the agreement with $W(\omega)/2$ and $\itof{}$ which was shifted
by the theoretical gap $\Delta_0 = 24\,$meV which is a prediction
of our calculations as, to our knowledge, no experimental data
exist for this material. The bottom frame of Fig.~\ref{fig:29}
presents our comparison between experimental and theoretical
optical scattering rates. The normal-state scattering rate
(gray lines) at $T=300\,$K gives evidence for the existence of
a high energy background as the experimental data (gray solid
line) are best fit by an MMP spin-fluctuation spectrum as
described by Eq.~(\ref{eq:63}) with $\omega_{SF} = 80\,$meV
and a high energy cutoff of $400\,$meV (gray dashed line).
The black lines compare the theoretical results (dashed line)
with experiment \cite{puchkov} (solid line) in the superconducting
state at
$T=10\,$K. The signature of the optical resonance, the sharp
rise in $\tau_{op}^{-1}(\omega)$ starting at around $50\,$meV
is correctly reproduced by theory. For $\omega > 120\,$meV the
experimental scattering rate shows only a weak energy dependence
and the theoretical prediction starts to deviate from experiment.
This is in contrast to our results found for all other compounds
and could be related to the fact that the Y124 compound shows
features of an underdoped system.

\section{Summary}

An extended Eliashberg theory can be applied to describe the
superconducting properties of hole doped high-$T_c$ cuprates.
The extension goes in two directions: first, it is essential to
allow the pairing potential to have $d_{x^2-y^2}$ symmetry, and,
second, the charge carrier-exchange boson interaction leading
to pairing has to be modeled using a phenomenological approach
because the microscopic origin of the attractive interaction
between the charge carriers is still unknown.

An anomalous steep rise in the superconducting state optical
scattering rate observed in optimally doped YBCO in the energy
range $50\le\omega\le90\,$meV was attributed to the coupling
of the charge carriers to an optical resonance located at about
$41\,$meV. This optical resonance has its counterpart in a
magnetic resonant mode which can be observed in YBCO at the
same energy by inelastic neutron scattering. This resonance
is not observed above $T_c$ in the normal-state and the
normal-state infrared scattering rate too does not develop any
anomaly. Further experimental data on the temperature dependence
of the infrared scattering rate and of the magnetic resonance
proved further agreement as the area under the optical resonance
is seen to have the same temperature dependence as the $41\,$meV
magnetic resonant mode. All this resulted in a definite procedure
which allows a phenomenological charge carrier-exchange
boson interaction spectral density $\itof{}$ to be derived
which reflects the
coupling of the charge carriers to the optical resonance and also
describes properly the almost linear frequency dependence of the
normal-state infrared scattering rate. Using this phenomenological
$\itof{}$ as the kernel of an extended Eliashberg theory allows us
not only to reproduce the experimental infrared optical data, it
also allows to reproduce properly the temperature dependence of the
microwave conductivity, of the London penetration depth, and
of numerous other superconducting properties.

This success justifies the extension of this analysis to other
compounds, like Bi2212, Tl2201, and Y124 for which less extensive
experimental data are available. The latest high quality optical data
on Bi2212 proved that, also in this case, the coupling of the charge
carriers to an optical resonance at $43\,$meV can be
associated with an anomalous steep rise in
the superconducting state infrared scattering
rate. In contrast to YBCO the anomaly in the infrared optical
scattering rate can also be observed in the normal state. This
optical resonance has, in the superconducting state, its
counterpart in a magnetic resonant mode observed by inelastic
neutron scattering at $43\,$meV. This mode has, so far, not been
observed in the normal state of optimally doped Bi2212. The
method to derive a phenomenological $\itof{}$ from optical data
developed for YBCO can also be applied to the compound Bi2212
and leads, again, to a temperature dependent kernel $\itof{}$
and the extended Eliashberg theory allows an excellent reproduction
of some superconducting state properties.

A similar anomaly can be observed in the superconducting state
low-temperature infrared scattering rate of Tl2201 and Y124. It
has, consequently, been interpreted as the coupling of charge
carriers to an optical resonance, not present above $T_c$.
Recently, in the monolayer compound Tl2201 a magnetic resonant
mode has been observed in the superconducting state but at a
slightly higher energy than predicted from optical data.
Nevertheless, this fact is quite important because it established
that the existence of a magnetic resonant mode is not restricted
to bilayer compounds. The existence of a magnetic resonant mode
is still to be proved in Y124, in which an optical resonance seems
to exist
at an energy of $38\,$meV. For both compounds, an extended
Eliashberg theory together with a phenomenologically derived
kernel $\itof{}$ resulted in a good reproduction of the optical
data. Not enough information about a possible temperature dependence
of the $\itof{}$ in these compounds is available to extend the
theoretical analysis to other superconducting state properties.

All this established a unified phenomenological picture for
hole doped high-$T_c$ cuprates which interpretes anomalies
in the charge carrier dynamics observed in optimally and
overdoped samples as a signature of spin degrees of freedom in
these compounds.

\section*{Acknowledgments}

This research was supported by the Natural Sciences and Engineering
Research Council of Canada (NSERC) and by the Canadian Institute
for Advanced Research (CIAR). The authors thank Drs. D.N. Basov,
C.C. Homes, and J.J. Tu for discussions and for making their data
available to us.

\end{document}